# Kaleidoscope Eyes: Microstructure and Optical Performance of Chiton Ocelli

Leanne Friedrich[1], Wai Sze Lam[2], Lyle Gordon[1], Paul Smeets[1], Robert Free[1], Lesley Brooker[3], Russell Chipman[2], and Derk Joester[1]*

[1] Materials Science and Engineering, Northwestern University, Evanston, IL 60208.
[2] College of Optical Sciences, University of Arizona, Tucson, AZ 85721.
[3] GeneCology Research Centre, University of the Sunshine Coast, Sippy Downs, QLD 4556, Australia.

* to whom correspondence should be addressed: d-joester@northwestern.edu

## Abstract

The chiton *Acanthopleura granulata* uses aragonitic lenses embedded in its shell to focus light onto photoreceptors. Because aragonite is biaxially birefringent, the microstructure of the lens greatly impacts the optical performance. In addition, the chiton lives in the intertidal, so lenses experience two environments with different refractive indices: air and water. Using EBSD, we find that the lens is polycrystalline and contains curved grain boundaries. A combination of large, twinned grains and nanotwins ensure that the aragonitic ⟨001⟩ axis is consistent across the lens. However, the orientation of the ⟨001⟩ axis relative to the shell varies between lenses. Ray tracing simulations predict the optical performance of lenses of various microstructures in wet and dry environments. Though twinning helps to limit birefringence-induced aberrations, variations in the orientation of the ⟨001⟩ axis between lenses lead to variations in focal lengths between lenses and cause image doubling in some lenses. As such, the birefringence of aragonite does not help the lens to transmit focused images in both air and water.

## 1 Introduction

Across a broad range of phyla, biominerals are used to reinforce tissues.[1] Functional roles further include homeostasis of essential ions such as $Ca^{2+}$, acceleration and gravity sensing, and orientation in magnetic fields.[1,2] Much less explored is the use of biominerals to manipulate light.[3–8] The extinct trilobites famously employed single-crystalline calcite lenses in their compound eyes, and the extant brittlestar *Ophiocoma wendtii* uses similar lenses for photoreception.[9–13] Both organisms evolved solutions to design problems that are general, such as the creation of ellipsoid lens volumes or the correction of spherical aberration, but also to the specific problem of double refraction of calcite. Double refraction is the consequence of birefringence, i.e. the dependence of the refractive index on the polarization and direction of propagation of light, and leads to the well-known doubling of images viewed through a calcite crystal.

While the choice of calcite as a material for lenses in an optical system may not be obvious, there are numerous examples where large, intricately shaped, and smoothly curving skeletal elements are comprised of one oriented single crystal of calcite and occluded macromolecules.[14] Creating the desired shape thus seems to be a manageable problem. Furthermore, birefringence in calcite is uniaxial, meaning that the material is optically isotropic for rays parallel to the axis of birefringence. With good control over non-equilibrium shape and lattice orientation of large single crystals, there are some straightforward ways to optimize lens design. Not so for aragonite, a polymorph of calcium carbonate frequently used by marine biomineralizing organisms. Aragonite not only displays biaxial birefringence, but also typically appears as a fine-grained polycrystalline material. Double refraction, scattering, and internal reflection at grain boundaries are expected to degrade the optical performance of polycrystalline lenses. Nevertheless, some chitons are uniquely known to construct an optical system from aragonite,[15,16] begging the question how they may have overcome the limitations inherent in the material they use.

Chitons are a diverse group of marine mollusks of the class Polyplacophora that have a characteristic dorsal shell comprised of eight plates or valves (**Figure 1**A). These valves are constructed from polycrystalline aragonite. Each valve is riddled with branching sensory and secretory channels connecting the dorsal surface to the mantle.[17] For eight genera of chitons, these channels terminate at the dorsal surface in light sensory organs called ocelli (Figure 1B-D).[15] For example, in *Acanthopleura granulata*, the West Indian fuzzy chiton, ocelli are composed of an aragonite lens suspended above the rhabdom, a 45-70 micron-deep cavity that contains roughly 180 sensory cells.[15,18] Lenses are biconvex and are covered by an aragonitic "cornea" (Figure 1C, D).[15,16,18] The sensory cavity and lens are flanked by aragonite containing pheomelanin pigment (Figure 1D).[19]

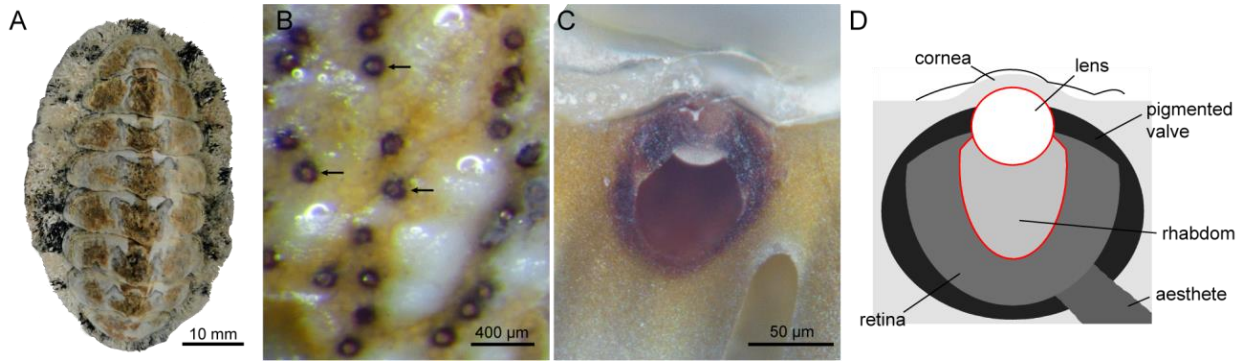

**Figure 1. A)** Dorsal aspect of *A. granulata*. **B)** Reflected light image of dorsal surface of valve wetted by water. Ocelli are recognizable by a ring of pigment (arrows). **C)** Reflected light image of cross-section of ocellus. Note that soft tissues have been removed and lens is partially obscured by pigmented valve. **D)** Schematic drawing of ocellus cross section.

Behavioral studies suggest that *A. granulata* may be capable of spatial vision.[15] Speiser and coworkers proposed that an aragonite lens may exhibit two focal lengths (for light of orthogonal polarization), thereby enabling chitons to perceive their environment while submerged in water and when exposed to air.[15] To test this hypothesis, a more detailed model of the lens is required. Herein, we investigate the microstructure of the *A. granulata* ocellus lens and evaluate its impact on the optical performance of ocelli using a combination of experiments and ray tracing simulations.

## 2  Results

### 2.1  Microstructure of the lens

In electron backscatter diffraction (EBSD) maps of ground and polished sections, lenses can be distinguished from the surrounding valve and the cornea by a roughly elliptical dark (non-indexed) line that circumscribes the lens **(Figure 2**A-C). We found that the average lens diameter was $d = 33 \pm 7.5$ µm ($N = 19$), and the maximum was $d = 41$ µm. Given the literature average of $55 \pm 2.3$ µm,[16] it is likely that the majority of the sections described herein did not contain the center of the lens.

Grain boundaries in the lens vary in shape. Many boundaries appear curved and interdigitated (Figure 2B, **Figure 3**B), while other sections present a fan-like morphology, with flat grain boundaries radiating from a central source (**Figure 3**A). In the valve, grains are sometimes equiaxed (Figure 3B) and sometimes columnar (Figure 2B).

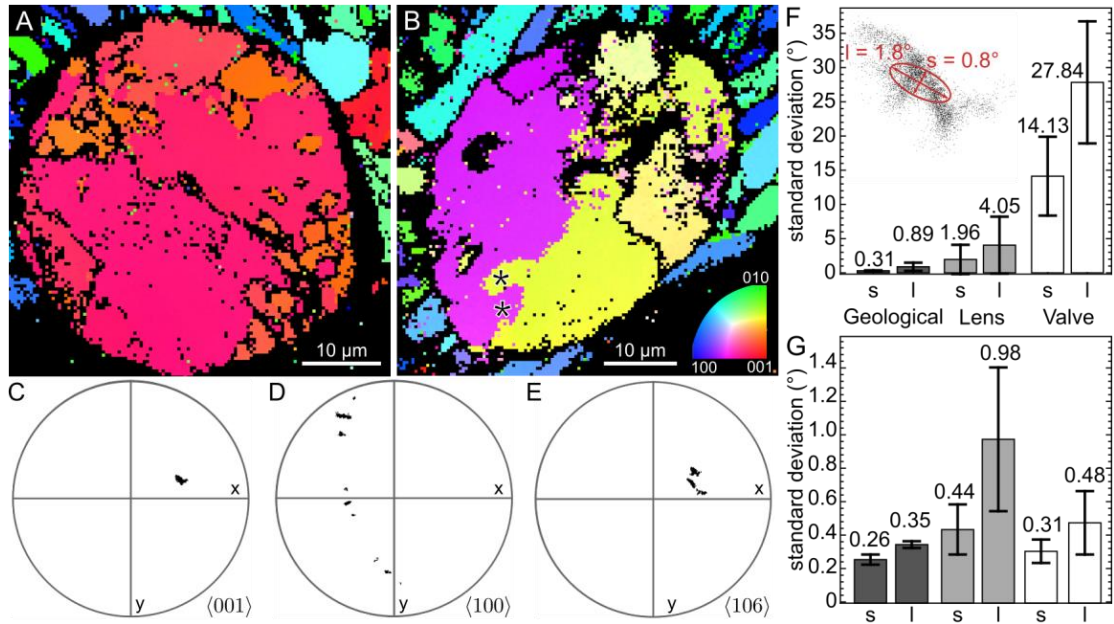

**Figure 2. A, B)** SEM-EBSD maps of embedded sections of lenses isolated from *A. granulata*. Orientation maps use an inverse pole figure color scheme, where the color indicates the crystallographic axis parallel to surface normal of the section (see inset in B). Areas rendered in black could not be indexed. Note curved, interdigitating grain boundaries (asterisks in B). **C,D,E)** Pole figures of the lens in (B). (E) shows the aragonitic optic axis. F,G) Width of the distribution of ⟨001⟩ orientations, in geological aragonite (N = 4 maps), within the lens (N = 18), and in the remainder of the valve (N = 1). The angular distribution was fit as a bivariate normal distribution along the long and short axes designated by the Kent distribution, where widths were measured separately for each map. **F)** Widths averaged over all grains. Inset: close-up of peak in (C), with long and short standard deviation annotated. **G)** Width averaged per grain. Grains are defined using a critical misorientation of 1° and a critical size of 100 pixels.

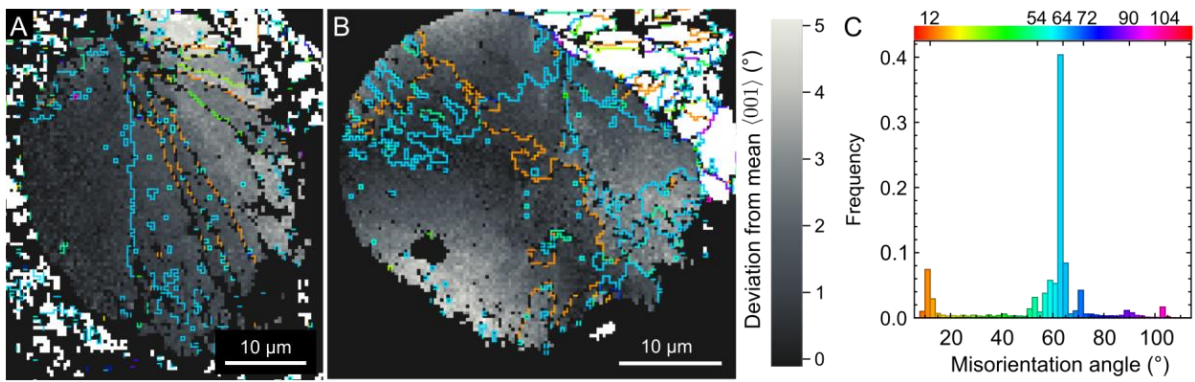

**Figure 3. A,B)** Abrupt and gradual changes of lattice orientation in the lens in plan (A) and cross section (B). Grain boundaries with misorientation angles above 10° are indicated by colored lines; the color indicates the angle of misorientation. Grains are shaded in gray levels according to the deviation of the local from the average orientation of the ⟨001⟩ axis in the lens. **C)** Histogram of misorientation angles for all lens grain boundaries. Bin width 2°.

The size of grains that fall into the lens varies considerably. Generally, there is a small number of grains that are significantly larger than average (Figure 2A,B). The mean grain diameter in the lens was 4.0 ± 4.2 μm compared to a median diameter of 2.6 μm. In some sections, lenses appear to be nearly single-crystalline (Figure 2A), with a single grain taking up 80% of the area. In others, the lens is composed of smaller grains (Figure 2B). On average, the largest grain comprises 30% ± 19% (*N*=18) of the lens area. In the remainder of the valve, grains are smaller, and the size distribution is narrower, with a mean of 2.6 ± 1.5 μm and a median of 2.0 μm (Figure S1).

## 2.2 Crystallographic texture

EBSD pole figures show dramatic differences in texture between the lens and the valve (Figure 2, Table S3, Table S4, Table S6, Table S5). In the lens, ⟨001⟩ axes are tightly aligned (Figure 2C,F). In contrast, ⟨100⟩ axes and the optic axis (approximately the ⟨106⟩) trace narrow arcs, indicating that only the ⟨001⟩ axis is aligned (Figure 2D,E). In the valve, the ⟨100⟩ and ⟨001⟩ axes both trace diffuse arcs, indicating that the valve microstructure has limited texture (Figure S4, Figure 2F). The orientation distribution of the ⟨001⟩ axis is seven times narrower in the lens than in the valve, but both regions still have wider ⟨001⟩ axis distributions than geological crystals (Figure 2F). Due to difficulties in preserving the cornea during sample preparation, we have relatively little information on its crystallographic texture (Figure S4). While ⟨001⟩ axes in the lens are highly aligned, we find that for any given lens, the angle between the optical axis of the lens and the mean ⟨001⟩ axis falls into the interval $\alpha = 21°$ - $82°$, seemingly at random.

In the lens, the orientation of the ⟨001⟩ axis changes gradually within individual grains (Figure 3B). Within individual grains, ⟨001⟩ axis distributions are wider in lens grains than in grains in the surrounding valve, likely because lens grains are larger (Figure 2G). ⟨001⟩ axis distributions in the small grains in the valve are the same size as distributions in the much larger grains of geological crystals (Figure 2G).

The axis and angle of misorientation, i.e. the parameters of a rotation that superposes the two lattices, describe orientation relationships between neighboring grains. The probability distribution of the misorientation angles exhibits a sharp global maximum at 64° (Figure 3C). The angle between the mean axis of misorientation of this type of grain boundary and the ⟨001⟩ axis is very small (0.06°), indicating that the ⟨001⟩ axis remains static across the grain boundary (Figure S2). Grain boundaries with a misorientation angle of 64° and misorientation axis of ⟨001⟩, are consistent with aragonitic twin boundaries on the {110} plane, although planar twinning on the {110} plane is not apparent at the microscale (Figure 3A,B). For local maxima at 12° and 54°, the angle between the mean axis of misorientation and the ⟨001⟩ axis is also rather small ($\leq 1.6°$). However, this angle is large enough to enable small changes in ⟨001⟩ axis orientation at 12° and 54° grain boundaries to accumulate into large changes in ⟨001⟩ axis orientation across the lens (Figure 3A). The axes of rotation for all other local maxima in the misorientation angle probability distribution do not appear to be aligned.

## 2.3 Nano-twinning

In addition to microscopic twinning, aragonite is known to display nano-twinning.[20–23] TEM imaging using bright field (BF) contrast indicates that nanotwins with a characteristic size of 10-100 nm are present in the lens of *A. granulata* (**Figure 4**). Nanotwins can be found within grains and at the interfaces between grains (Figure 4A) and have coherent (110) interfaces (Figure 4B,C). Within the nanotwin, spot doubling is apparent in fast Fourier transforms of the lattice. Superposition of matrix and twin plane reflections produces spot doubling.[21]

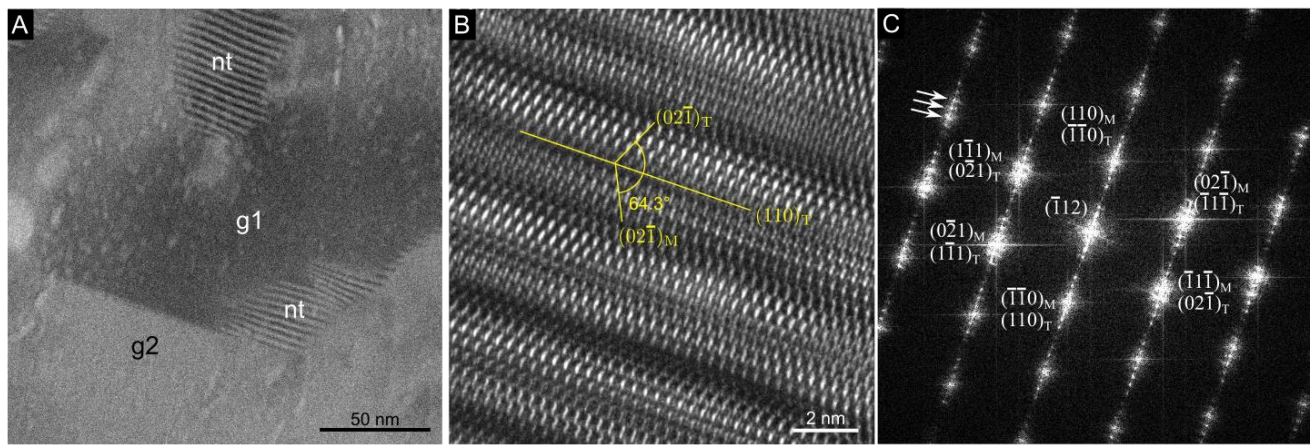

**Figure 4. A)** Bright field TEM image of nanotwins (nt) at the interface between two grains (g1, g2). **B)** HRTEM image of a nanotwin on the [$\bar{1}12$] zone axis. Matrix (M) and twin (T) planes are annotated. **C)** Fast Fourier transform of (B), showing spot doubling (arrows) which is indicative of twinning.

## 2.4 Differential solubility

When ground and polished sections of lenses were exposed to slightly acidic, ultrapure water, small amounts of aragonite dissolved within minutes. When inspected by SEM, etched surfaces share several features. Most prominent is a set of thin, approximately elliptical grooves that resemble eccentric growth bands, i.e. the growth bands one would expect if the growth rate were dependent on the direction (**Figure 5**). The grooves do not necessarily superimpose on grain boundaries, but they run perpendicular to grain boundaries in sections where a fan-shaped microstructure is apparent (Figure 5). Etch patterns follow a core-periphery structure. In the core, i.e. at the center of the growth bands, etching produces deep etch pits with no dominant orientation (Figure S5G,J). In the periphery, deep, faceted etch pits (Figure S5H,K) and shallow, curved etch pits (Figure S5I,L) run radially away from the core.

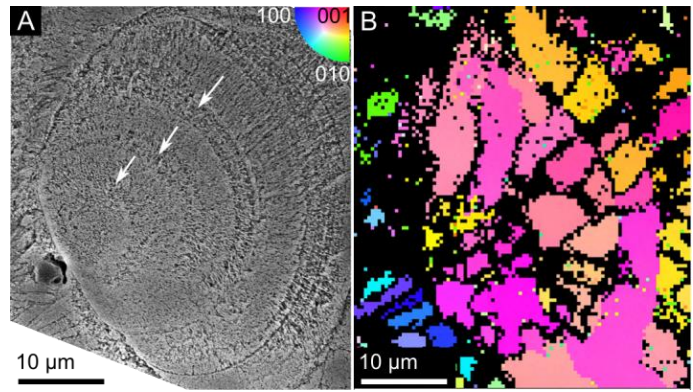

**Figure 5. A**) SEM image of lightly etched lens cross-section. Note eccentric growth bands (white arrows). **B**) EBSD grain orientation map of the same lens as in (A), using inverse pole figure coloring.

## 2.5 Simulations

Clearly, there is a significant number of twin and other grain boundaries in the lens of *A. granulata*. This is important for its optical performance. When a light ray enters a birefringent medium across an interface, it splits into a 'high' ray that propagates in a direction of high refractive index, and a 'low' ray that propagates in a direction of low refractive index. In a biaxial birefringent medium such as aragonite, the directions of propagation of rays depend on the relative and absolute orientation of the grains the rays encounter on their path through the lens. The focal position, focal region size (longitudinal and transverse aberrations), and astigmatism thus all depend on the microstructure. To investigate the impact of microstructure on the optical performance of the ocelli, we conducted ray-tracing simulations using Polaris ray tracing software.[24] To identify design rules, and to reduce computational complexity, we idealized the lens structure in the following way (**Figure 6**):

1. Relevant interfaces were modeled as hemi-ellipsoids and planes.
2. The cornea was modeled as an *isotropic* material with the average refractive index of aragonite (n=1.632).
3. Unless otherwise noted, the lens was modeled as a single crystal.
4. Incident light was modeled as monochromatic ($\lambda = 500$ nm).
5. Paraxial, intermediate, and peripheral rays were generated parallel to the principal ray. While most rays passed through the lens and entered the rhabdom cavity, some peripheral rays experienced total internal reflection and were not further considered.

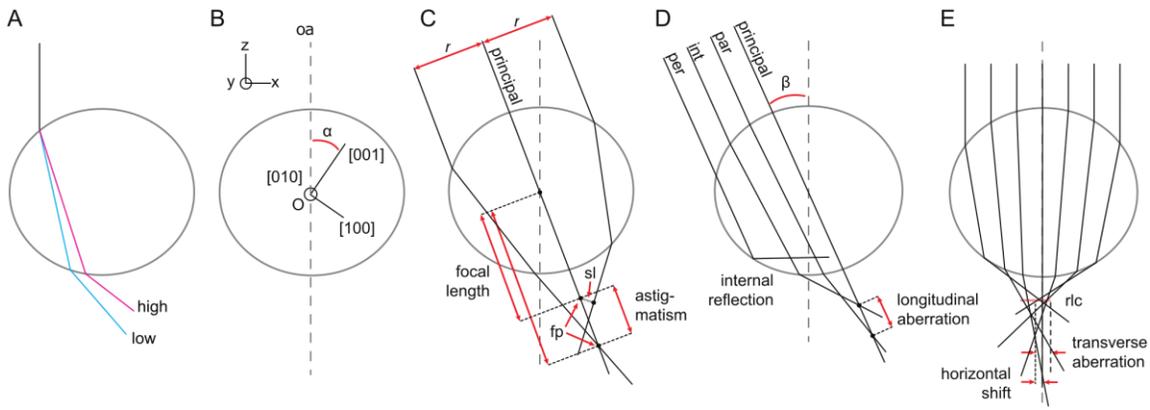

**Figure 6.** Ray tracing conventions. **A**) Crossing the air-aragonite interface, a light ray splits into a 'high' and a 'low' ray. **B**) The center of the lens is the origin of the lab coordinate system. The lab y-axis is parallel to the aragonite [010]-axis, and the lens optical axis (oa) (which is defined by lens geometry and is different from the aragonitic optic axis) is parallel to the lab z-axis. A rotation by angle α about the aragonite [010]/lab y-axis superposes the optical axis and the aragonite [001] axis. **C**) The principal ray passes through the origin (O). The focal length for any ray parallel to the principal ray at some distance r is the distance from the origin of the focal point (fp). Focal points of intersecting rays are the point of intersection. Focal points for non-intersecting rays are the point of intersection of the principal ray with the shortest line (sl) between the two rays. The range of focal lengths for all parallel rays at some distance r from the principal ray is the astigmatism. **D**) A rotation by the angle of incidence, β, about the lab y-axis superposes the principal ray and the optical axis. Sets of paraxial (par), intermediate (int), and peripheral (per) rays were generated parallel to the principal ray, at increasing radial distance. Some peripheral rays display total internal reflection. The total range of focal lengths for all rays is the longitudinal aberration. **E**) The region of least confusion (rlc) is the area where rays collectively have the smallest root mean square distance from the principal ray. Transverse aberration is the diameter of the region of least confusion, and horizontal shift is the distance of the centroid of the rlc from the optical axis.

### 2.5.1 Focal length

After passing through the lens, paraxial, intermediate, and peripheral rays are deflected towards, but do not necessarily intersect the principal ray. Focal points for each ray were therefore determined as the intersection of the principal ray with the shortest line segment between the ray and the principal ray (Figure 6C). Unless otherwise noted, the focal length for a given set of rays was calculated as the mean distance of all focal points from the origin.

Ignoring spherical aberration, an isotropic lens has a single focal surface, and photoreceptors placed on the focal surface provide the sharpest image. In contrast, birefringent lenses produce one focal surface each for the high and the low rays. To begin, consider only the center point of the focal surface, i.e. the focal point of the paraxial rays where $\beta = 0°$. Ray tracing reveals that focal lengths for high and low paraxial rays are within ~0.5 µm of each other if $\alpha = 0°$. With increasing α, the focal length of the high rays does not change, but that of low rays almost doubles as α approaches 90˚ (**Figure 7**). Thus, the high ray image is always in focus in the same plane, inside the rhabdom. When $\alpha = 0°$, the low ray image is superimposed on the high ray image, but as α increases, the low ray image shifts lower in the rhabdom, exiting the rhabdom between $\alpha = 60°$ and $\alpha = 90°$.

Next, consider the edges of the focal surface, determined by the focal length of the paraxial rays where $\beta > 0°$. As β increases, the focal length of the high paraxial rays remains constant, while the focal length of the low paraxial rays increases (**Figure 8A**). Consequently, a lens with $\alpha = 0°$ has two focal surfaces which intersect at their centers but diverge at their peripheries.

Finally, consider the environment in which the chiton resides. Rays entering the cornea from air are refracted more strongly than rays entering from water because air has a lower refractive index than water. As a result, lenses in air have a shorter focal length than lenses in water (Figure 7A,E). However, unlike in isotropic lenses, in birefringent lenses the focal length ratio between air and water is not fixed (Figure S6). This ratio varies between points at equivalent radii and for varying ⟨001⟩ orientations.

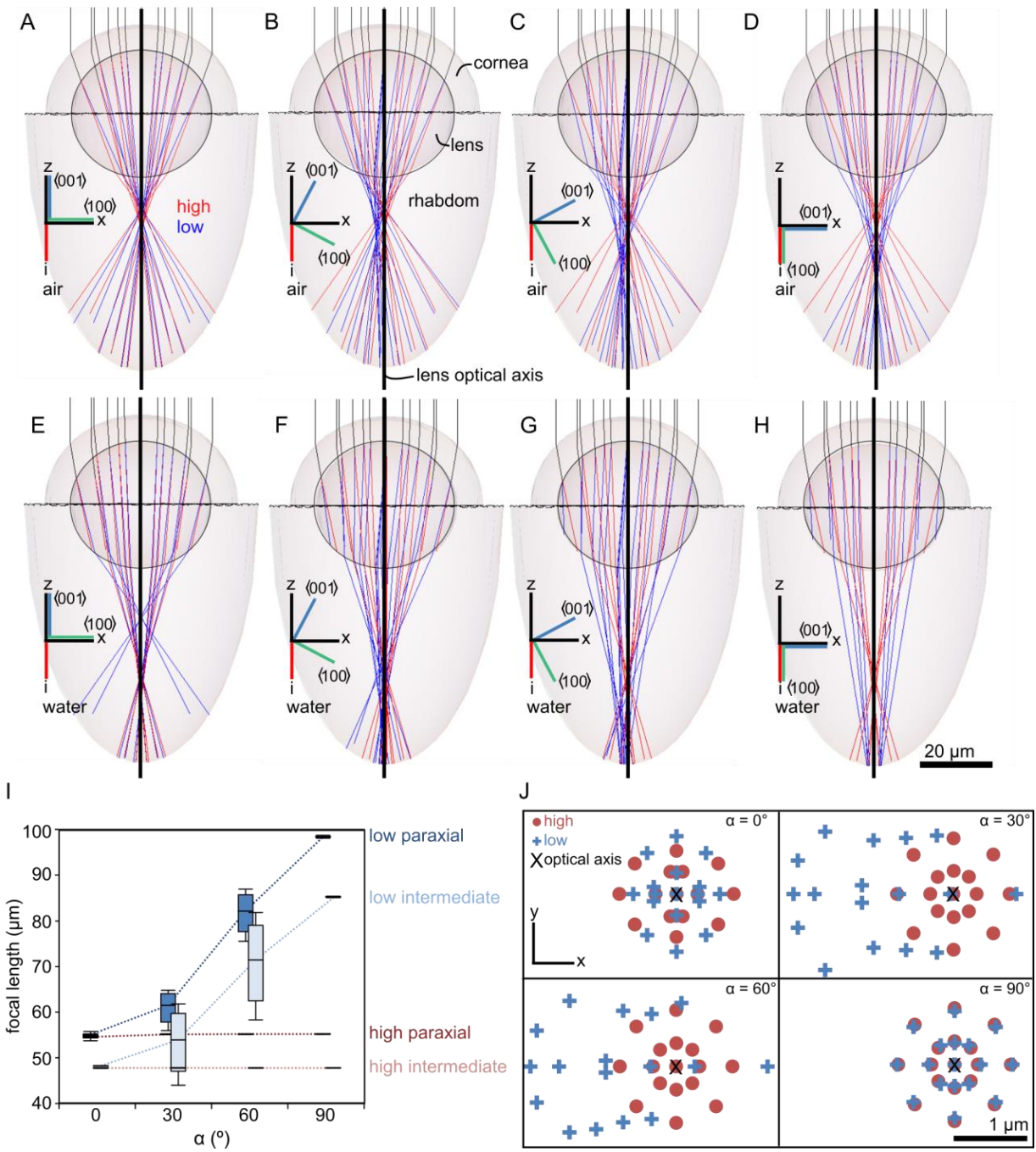

**Figure 7. A-H)** All simulated rays, in the ⟨010⟩ orthographic projection. **I)** Focal lengths in water. Boxes and whiskers indicate quartiles. **J)** Intersections between paraxial and intermediate rays and the plane of least confusion in water, demonstrating transverse aberration and horizontal displacement.

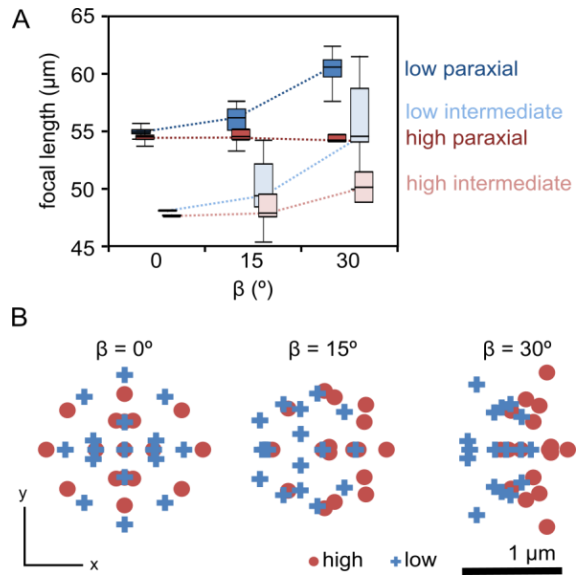

**Figure 8. A)** Focal lengths where $\alpha = 0°$. **B)** Orthographic projection of intersections between rays and the plane of least confusion, demonstrating transverse aberration. Horizontal displacement is not illustrated because the principal ray is deflected at high $\beta$.

### 2.5.2 Horizontal shift

In isotropic lenses, all rays remain centered around the optical axis. However, in birefringent lenses, focal points can displace horizontally relative to the center of the incident ray of light, which causes image doubling. Horizontal shift was calculated as the distance between the focal points and the central axis of the rhabdom (Figure 6E). At $\alpha = 0°$ and $\alpha = 90°$, the focal point is aligned with the lens optical axis. However, at intermediate $\alpha$ values, the focal point shifts horizontally along the crystallographic $\langle 100 \rangle$ direction (Figure 7B, C, F, G, J).

### 2.5.3 Astigmatism

In isotropic lenses, rotational asymmetry in the lens geometry causes astigmatism, where rays that enter the lens at equivalent distances from the lens optical axis have different focal lengths. In birefringent lenses, rotational asymmetry in the crystal structure can also cause astigmatism.

Astigmatism was calculated as the difference between the maximum and minimum in focal length among rays originating the same distance from the lens optical axis (Figure 6C). Consider paraxial rays entering the lens at $\beta = 0°$. At $\alpha = 0°$ and $\alpha = 90°$, high and low paraxial rays experience no astigmatism. However, at intermediate $\alpha$ values, paraxial low rays inhabit a range of focal lengths. Intermediate low rays experience even larger astigmatism values, up to 11 µm in water (Figure 7I). Similarly, increasing the angle of incidence increases astigmatism. For a lens with $\alpha = 0°$, increasing $\beta$ introduces astigmatism not only among the low rays, but also among high rays (Figure 8).

### 2.5.4 Longitudinal aberration

All lenses with spherical surfaces experience longitudinal spherical aberration, in which rays at the edge of the lens have a shorter focal length than rays at the center of the lens. In birefringent lenses, this geometric aberration still holds. Longitudinal aberration was calculated as the distance between the shortest paraxial and the longest intermediate focal length (Figure 6D). Where $\beta = 0°$, at all $\alpha$ values the longitudinal aberration of the high rays was ~7 µm in water. At $\alpha = 0°$ and $\alpha = 90°$, the longitudinal aberration of the low beam is comparable to that of the high beam. However, at intermediate $\alpha$ values, longitudinal aberration of the low beam is large, mainly because of astigmatism. The difference between the median paraxial and intermediate rays is not much larger for low rays than high rays, indicating that this longitudinal aberration comes largely from geometry, not birefringence.

### 2.5.5 Transverse aberration

Longitudinal aberration, astigmatism, and horizontal shift blur and distort images. As a result, light originating from a point source focuses to a region of least confusion instead of a point. A detector placed at the position of the region of least confusion will therefore record the least blurry image. The size of the region of least confusion is defined as the size of the ellipse which encompasses the intersections of all rays with the plane of the region of least confusion (Figure 6E). The lengths of the ellipse half-axes indicate transverse aberration, and the difference between those lengths is a supplemental metric of astigmatism. Larger, less symmetric regions of least confusion produce low image quality.

In an isotropic lens, the region of least confusion is circular for lenses without astigmatism and elliptical for lenses with astigmatism. In this birefringent lens, the region of least confusion can exhibit a more complex shape. For $\beta = 0°$, high rays exhibit a circular region of least confusion when $\alpha > 0°$ and a slightly elliptical region when $\alpha = 0°$, while low rays exhibit an elliptical region of least confusion at $\alpha = 0°$ and $\alpha = 90°$ (Figure 7J). At intermediate $\alpha$ values, low rays focus to a larger, asymmetric region of least confusion (Figure 7J). Asymmetry becomes even more pronounced when $\beta > 0°$ (Figure 8B).

## 3 Discussion

### 3.1 Lens microstructure and growth

The microstructure of the lens suggests a possible growth route and has implications for the ability of the lens to transmit images. EBSD indicates that the lens contains a core region with small, equiaxed grains, and a periphery containing larger, fan-shaped grains with grain boundaries that have characteristics of twinning. Aragonitic twinning occurs most commonly on {110} planes, with a ⟨001⟩ misorientation axis and 63.75˚ misorientation angle.[25] Twinning is commonly observed in geologic and other biogenic aragonites such as fish otoliths, bivalve shells, and coral skeletons.[21,26–31] It is well documented in mollusks, including in crossed-lamellar structures such as the mesostracum and myostracum of the chiton valve,[20,21,27,32] helical fibers in the pteropod *Cuvierina*,[33] prismatic aragonite structures in bivalves and gastropods,[29,30,34] and the prismatic tegmentum in which chiton ocelli reside. In polycyclic twins of aragonite, the angle between grains separated by two or three twin boundaries can be either 11.25˚ or 52.5˚. It is therefore possible that the 12° and 54° grain boundaries observed in EBSD result from cyclic twinning in the lens core, as has been observed in foliated aragonite in other mollusks.[29–31,35] Most polycyclic, polysynthetic, and deformation twins in geological and biogenic aragonite are coherent and exhibit planar grain boundaries.[36,37] In contrast, twin boundaries in ocelli frequently exhibit micron-scale curvatures.

The nanotwins observed using TEM may limit incoherency strain at some curved grain boundaries. Further, nanotwins may constitute a toughening mechanism which compensates for the loss in strength caused by large grain sizes in the lens.[22] Nanotwins could be the cause of poor EBSD indexing in regions including the core of the lens, the lens-valve interface, and some grain boundaries. Nanotwins or strain may also be responsible for gradual changes in ⟨001⟩ axis orientation within grains. These gradual changes in orientation across the ocelli lens are qualitatively similar but smaller than the gradual changes in the calcite ⟨0001⟩ axis across trilobite compound eye lenses.[38] Because grain shapes and sizes vary between lens cross-sections, the absolute ⟨001⟩ orientation varies between lens, and the lens contains nanotwins that are not detectable with EBSD, it is necessary to characterize many cross-sections of many lenses on multiple length scales in order to fully understand the complex three-dimensional microstructures of these lenses.

Etching reveals growth bands which run perpendicular to the fan-shaped microstructure observed in EBSD. We expect the etch rate to increase wherever impurities, other structural defects, and/or strain increase the local solubility. When crystals grow on a curved interface, fluctuations in the growth rate frequently result in the incorporation of impurities as growth bands that provide a record of the shape of the interface.[39] Impurities also frequently segregate to grain boundaries and modulate etch rates parallel to the grain boundaries. In fact, some grain boundaries seem to etch more rapidly. Together, EBSD, TEM, and etching illustrate a possible growth rate for the lens. The lens may nucleate quickly as a finely twinned core, and then large fan-shaped grains grow off the core, maintaining a curved growth surface. Nanotwins may alleviate incoherency strain associated with the curved grain boundaries which result from this growth route.

## 3.2 Optical implications of microstructure

While estimating the lens optical axis from optical images introduces some uncertainty, the range of ⟨001⟩ orientations observed with EBSD indicates that the organism does not align the ⟨001⟩ with the lens optical axis, which is consistent with a previous study.[40] The orientation of the ⟨001⟩ axis has implications for the chiton's ability to see in both air and water. *A. granulata* live high in the intertidal zone and are exposed to air for long periods of time. Speiser and coworkers predicted that the birefringence of the aragonitic lens might allow the chiton to form an in-focus image of their environment both under water and in air.[15] At $\alpha = 90°$, the low ray focal length in air is close to the high ray focal length in water, so if sensory cells were placed on that plane, the retina would collect a clear image in both air and water. However, because all ocelli we examined have $\alpha < 90°$, the positions of the high and low focal planes vary between air and water, so the birefringent lens is no more useful than an isotropic lens for adapting to a tidal environment. Filling the entire cavity with sensory cells is thus useful because of the large variability in focal lengths among α values and between air and water.

Simulations indicate that a lens with a crystallographic ⟨001⟩ axis that is not aligned with the lens optical axis could be advantageous. At $\alpha = 0°$, the high and low focal surfaces overlap at the center. However, because the focal surfaces are ~7 µm apart at the image periphery, which is equal to the photoreceptor size, imaging both the high and low focal surfaces simultaneously would result in image doubling at the image periphery. The organism could only separate the two sets of rays and limit aberrations if its photoreceptors were sensitive to polarization. However, if the chiton has spatial sensitivity but not polarization sensitivity, a lens with $\alpha > 0°$ would be advantageous because the high and low focal planes would be spatially separated, allowing the chiton to limit image doubling through spatial filtering. Still, the high ray image would be blurred by out-of-focus contributions from the low rays, and vice versa. Furthermore, if the chiton does not have the sensory capabilities to spatially filter focal surfaces, minimizing α would limit the blurring and doubling produced by birefringence.

Twinning is beneficial because of the dependence of focal length on crystallographic orientation. In lenses with a varying ⟨001⟩ orientation, low ray images from different directions fall on varying focal planes. In a twinned lens, the ⟨001⟩ axis remains constant, which ensures that all low rays converge on a single focal surface. However, because the ⟨100⟩ and ⟨010⟩ axes of the grains in the lens are not aligned, and lenses have intermediate α-values, horizontal displacement will cause image doubling across each grain boundary in ocelli. Doubling at each grain boundary could cause each lens to transmit several distinct images, which is visible in images projected through the lens (Figure S9). We did not simulate nanotwins, but because the nanotwins are smaller than the wavelength of light, and horizontal displacements would be small in nanoscale grains, we do not expect nanotwins to impact the optical properties of the lens.

Behavioral studies have indicated that chitons have the capability for spatial vision, and simulations indicate that the spatial resolution of images transmitted through individual lenses is suited to the retina.[15] Pigmented apertures are wider than 14 µm, but many rays outside of the 14 µm diameter experience total internal reflection and do not influence the quality of images that reach the rhabdom. The ellipses of least confusion for rays generated less than 14 µm from the principal ray are all smaller than 3 µm, so even the largest regions of least confusion measured in this study are smaller than the photoreceptor size of 7 µm. Though asymmetric regions of least confusion are reminiscent of comatic aberrations and astigmatism, and they degrade the image quality, the chiton does not have the sensory spatial resolution to detect such aberrations.

## 4 Conclusion

Microstructural analysis and simulations indicate that polycrystalline ocelli lenses are similar in optical properties to single-crystalline lenses, which allow them to transmit useful images. While lenses exhibit diverse grain morphology, and light reaches a different number of interfaces in each lens, all lenses we examined have large grains which limit the number of interfaces that light crosses. This is beneficial for maximizing transmittance and minimizing deflection, allowing more focused signals to reach the rhabdom. Image scattering is limited because the lens is populated with twin boundaries rather than randomly oriented grain boundaries.

Twinning may allow the lens to form quickly without sacrificing image quality.[28] Because a lens composed of two twins has the same focal length and transmission as a single crystal lens, the lens can be composed of many twins. If half of the rays entering the lens pass through a twin boundary, but the other half only pass through one grain, all rays can still be resolved together, as if they all passed through a single crystal lens.

Behavioral studies have shown that chitons are capable of spatial imaging,[15] and this study shows that lenses are capable of transmitting images, but the chiton's sensory capabilities to process images from the ocelli remain largely undocumented. The ocellus rhabdom compensates for the variability of the lens microstructure. It is impossible to anticipate the focal length of the lens without knowing its microstructure. Thus, it is appropriate that ocelli sensory cells fill the entire cavity below the lens. This allows the chiton to gather signal from anywhere in the rhabdom, so it can adjust to the diversity of lens microstructures.

The lens grain structure limits image scattering through consistency of the ⟨001⟩ axis, so the main limitations in image quality are longitudinal aberration, transverse aberration and birefringence. Trilobites and brittlestars both have mechanisms to compensate for aberrations that result from the lens geometry, through variations in refractive index and lens shape.[9–11] While the shape of chiton ocelli lenses has been heavily studied before, little is known about the composition of the lenses. If the refractive index varies throughout the lens through impurities, longitudinal and transverse aberration could be limited. The chiton could also use the rhabdom shape to adjust to the crystallography of the lens. Aesthetes run at an angle to the shell surface rather than traveling directly upwards, so the rhabdom angle could help the chiton adjust to the tilted ⟨001⟩ axis.[41–43] Whether the chiton employs either of these techniques is still unknown. From this study, it is clear that the *A. granulata* ocellus possesses a polycrystalline, twinned lens which is capable of transmitting useful images.

# 5 Experimental section

## 5.1 Preparation of sections

For samples which were etched after EBSD, dried *A. granulata* shells collected in the Florida Keys were purchased from a professional collector (Shellmama's Quality Shells). For samples which were etched before EBSD, dried *A. granulata* shells collected in Venezuela were obtained from the University of Alabama. Valves were excised using tweezers and broken into smaller segments using a mortar and pestle. Sections were prepared from a) valve segments with and without lenses, and b) whole lenses extracted from valves using a razor blade. Samples were embedded in epoxy (Epo-Tek 301) and polymerized overnight at room temperature. Sections were performed such that the plane of section was approximately normal to the surface of the valve (cross section) or such that the plane normal was parallel to the optical axis (plan section). Lens sections were prepared by sequential grinding using SiC paper (600, 800, and 1200 grit). Ground sections were then sequentially polished using polycrystalline aqueous diamond (3 µm and 1 µm), followed by alumina (0.05 µm) polishing suspensions. Polished specimens were secured to aluminum scanning electron microscope (SEM) stubs using cyanoacrylate adhesive.

Orientation of cross and plan sections was confirmed by SEM. Sections were confirmed as "plan" if the lens was surrounded by pigment, with micro-aesthetes running normal to the polished surface. Sections were confirmed as "cross" if there was a smoothly curved lower lens surface, pigment at the edge of the lower lens surface, and, in some sections, a cornea present. All other sections were categorized as oblique. For plan sections, the orientation of the optical axis was assumed to be normal to the plane. For cross sections, the direction of the optical axis was approximated from optical micrographs and assumed to be in the plane.

Reference samples were prepared from pieces cut from a geological aragonite crystal, embedded with either the ⟨001⟩ or ⟨100⟩ oriented approximately normal to the sample surface, ground, and polished as described above. The final cross-sectional area of the samples was approximately 2 mm$^2$.

## 5.2 Electron Backscatter Diffraction (EBSD)

Ground and polished sections of 24 lenses, one shell section, and sections of geological aragonite were examined using electron backscatter diffraction (EBSD). In seven samples, the plane of the section was approximately perpendicular to the surface of the valve; we refer to these as cross-sections. In fifteen samples the plane was approximately parallel to the surface of the shell (plan sections), and in two cases it was oblique.

Uncoated samples were mounted on a 70º pre-tilted SEM sample holder and observed in a FEI Quanta 600F Environmental Field Emission SEM operated at a water vapor partial pressure of 0.9 Torr, an accelerating voltage of 30 keV, and a working distance of 10 mm. Kikuchi patterns were collected from ground and polished biological specimens (step size 0.3 µm-1.1 µm) and reference samples (step size 17 µm - 24 µm) using an Oxford Nordllys II detector

## 5.3 Analysis of EBSD Data

EBSD data was processed using the Oxford AZtecHKL package. Misorientation between neighboring grains was determined using a 1° critical misorientation and orthorhombic symmetry operators.[44–46] Only grains at least ten pixels in size were considered for grain size analysis.[47,48]

Computations for statistical analyses of grain orientation were performed using Wolfram Mathematica 10, in part using code adapted from Leong and Carlile's Spak.[49] The mean orientation of a given crystallographic direction was determined by fitting the 5-parameter Kent (Fisher-Bingham) distribution. Only grains at least 2 pixels in size were considered in this analysis. In addition to the mean orientation, the most and least dense directions of the distribution, the standard deviations of the distribution along those directions, the concentration $\kappa$ (large value indicates tight alignment), and the ellipticity $\beta$ (large values indicate rotational asymmetry) were determined.

## 5.4 Etching

For etching, ground and polished samples were submerged in water (~400 mL, pH = 5.5) and agitated on a rocking table at 30 rpm for 15 minutes. For some samples, EBSD orientation maps were determined before etching. Samples were dried with compressed air, mounted on aluminum SEM stubs using cyanoacrylate adhesive, coated with 10 nm of platinum using a Denton Desk III sputter coater and grounded using colloidal silver paint. Etched, coated samples were observed in a Hitachi S-3400N-II SEM operated at an accelerating voltage of 20 keV, a probe current of 50 µA and a working distance of 10 mm. Additional coated, etched samples were observed in a Hitachi S4800-II cFEG SEM at an accelerating voltage of 15 keV, a probe current of 10 µA and a working distance of 5 mm.

## 5.5 Focused Ion Beam (FIB) Preparation of TEM lamellae

TEM lamellae were prepared from ground and polished cross-sections of epoxy-embedded ocelli. As judged by EBSD data, several grain boundaries with different orientations of the lens were targeted for the liftout procedure. Specifically, a dual-beam FIB/SEM (FEI Helios Nanolab) with a gallium liquid metal ion source (LMIS) operating at an accelerating voltage of 2–30 kV was used to prepare FIB samples for TEM. A layer of protective platinum (~300 nm) was deposited on a 2 µm x 12 µm area of interest by using the electron beam (5 kV, 1.4 nA) through decomposition of a (methylcyclopentadienyl)-trimethyl platinum precursor gas (FEI Helios Nanolab). Additionally, ~1 µm thick protective platinum layer was deposited on the same area using the ion beam (30 kV, 93 pA) through decomposition of the same precursor gas. Subsequently, two trenches were cut to allow for a roughlly 1.5 µm thick lamella of the lens. Next, the micromanipulator was welded onto the lamellae, and the sample was cut loose from the bulk material. An *in situ* liftout of the sample was performed, and the lamellae was welded onto a copper TEM half-grid. After thinning to about 100 nm in a sub-region of the lamellae (5 kV, 81 pA), the thin section was cleaned at increased angles with respect to the ion-milled surface using low voltage and current (2 kV, 28 pA), to remove amorphous material resulting from higher current milling. The final sections were roughly 90 nm thick, as determined from the electron beam imaging in the SEM.

## 5.6 HRTEM and STEM Imaging

Bright Field scanning TEM (BF-STEM) images of FIB-prepared lamellae were acquired on a Hitachi HD-2300 STEM using a room temperature single tilt side-entry holder (Hitachi). Specifically, the STEM was operated at an accelerating voltage 200 kV, using a probe current of 168 pA. 2048 × 2048 pixel images were acquired using a dwell time of 2 µs and a pixel size of 1.4 × 1.4 Å, yielding a total electron dose of 1071 e$^-$ Å$^{-2}$.

High resolution TEM (HRTEM) images of FIB-prepared lamellae were acquired on a JEOL GrandARM 300F, using a side-entry room temperature double tilt sample holder (Gatan). Specifically, the TEM was operated at an accelerating voltage of 300 kV and a beam current of 10 µA. The electron dose of the acquired images on the sample was calculated from the transmitted intensity to approximate ~4.8 × 10$^3$ e$^-$ Å$^{-2}$ (4096 × 4096 pixels; pixel size 0.11 × 0.11 Å).

## 5.7 Simulations

For ray tracing simulations, we defined the following interfaces: environment-cornea, cornea-lens, lens-rhabdom, and rhabdom-shell. The former three were modeled as oblate, the latter as prolate hemi-spheroids centered on the origin, using the Cartesian equation (1). Note that in in *A. granulata,* lenses can take spheroidal or triaxial ellipsoidal shape;[16] we chose to ignore the latter for simplicity.

$$\frac{x^2 + y^2}{a^2} + \frac{z^2}{c^2} = 1 \qquad (1)$$

The positive lab *z*-axis is identical to the surface normal of the shell and the optical axis (axis of symmetry) of the lens. Note that the lens optical axis is distinct from the aragonite optic axis. The half-axes *a* and *c* were determined from optical images of embedded cross sections (**Table 1**, Figure 6).

| Table 1. Geometric Parameters. | | | |
|---|---|---|---|
| Interface | $a$ [µm] | $c$ [µm] | half space |
| environment-cornea | 27.3 | 25.3 | $z>0$ |
| cornea-lens | 20.0 | 18.0 | $z>0$ |
| lens-rhabdom | 20.0 | 18.0 | $z<0$ |
| rhabdom-shell | 33.0 | 73.0 | $z<0$ |

An incident principal ray through the origin, and thus the center of the lens, was generated in the environment above the lens. In some simulations, the incident principal ray was rotated counter-clockwise around the positive lab *y*-axis by an angle β (Figure 6). Eight more parallel rays in azimuthal increments of 45° were generated in each of three rings surrounding the principal ray at a distance of 6.7 µm (paraxial), 13.3 µm (intermediate), and 20 µm (peripheral). All rays were modeled as collimated, unpolarized light with a wavelength $\lambda = 500$ nm.

For ray tracing, the environment was modeled as air ($n = 1$) or water ($n = 1.33$); the cornea as an isotropic material with the average refractive index of aragonite ($n = 1.632$); the lens as a single crystal of aragonite ($n_\alpha = 1.530$, $n_\beta = 1.680$, $n_\gamma = 1.686$); and the rhabdom was modeled as water. The orientation of the aragonite lattice in the lens was such that $[100]_{ar}\|[100]_{lab}$, $[010]_{ar}\|[010]_{lab}$, and $[001]_{ar}\|[001]_{lab}$. In some experiments, the lattice was rotated clockwise around the positive [010] direction (lab *y*-axis) by an angle α (Figure 6). In some simulations, an incoherent twin interface parallel to the lab *z*-plane and through the origin was added, separating the lens into an upper segment with a lattice orientation identical to the one described above, and a lower segment in which the aragonite lattice was rotated 64° counterclockwise about the positive [001] direction. A non-twin grain boundary was simulated by rotating the [001] direction by 64° about the [010] direction in the lower segment.

Aragonite exhibits biaxial birefringence. Thus, any ray propagating in aragonite has one of two eigenstates, which are a combination of polarization and refractive index. The ray with 'high' eigenstate has the higher of the two possible refractive indices for a given direction and is therefore also referred to as the 'slow ray'. A ray in a low eigenstate experiences a lower refractive index and is also referred to as the 'fast ray'. Here, we use the terms 'high' and 'low' to indicate the eigenstate of rays. Rays passing from an isotropic material into aragonite are each split into a high and a low ray that propagate in different directions (Figure 6B). Because the high ray experiences a higher refractive index, it is refracted more strongly than the low ray.

Depending on the angle of incidence at any given interface, rays may be partially or totally internally reflected (Figure 6C). Calculating the direction of propagation, polarization, and intensity of rays is not trivial; we used the Polaris engine to do so.[24] Intensities of transmitted beams were calculated assuming that the initial rays are not polarized.

# 6 Acknowledgements

The study resulting in this publication was assisted by a grant from the Undergraduate Research Grant Program which is administered by Northwestern University's Office of the Provost. However, the conclusions, opinions, and other statements in this publication are the author's and not necessarily those of the sponsoring institution. This research was sponsored by the Chemistry of Life Processes Institute CAURS Fellowship.

This work made use of the EPIC facility (NUANCE Center-Northwestern University), which has received support from the MRSEC program (NSF DMR-1121262) at the Materials Research Center, The Nanoscale Science and Engineering Center (EEC-0118025/003), both programs of the National Science Foundation; the State of Illinois; and Northwestern University. This work also made use of the OMM Facility supported by the MRSEC program of the National Science Foundation (DMR-1121262) at the Materials Research Center of Northwestern University.

We thank Alberto Pérez-Huerta (University of Alabama) for providing shell samples.

# 8 TOC Entry

**In a manner observed in few other organisms, chitons employ a crystalline lens for vision.** Microstructural analyses and ray tracing simulations indicate that *A. granulata* possesses a unique aragonitic lens that offers the chiton the potential for spatial imaging. Twinning and large grains mitigate aberrations and image doubling caused by the birefringence of aragonite.

Keywords: chiton, aragonite, biomineralization, birefringence, twinning

Leanne Friedrich[1], Wai Sze Lam[2], Lyle Gordon[1], Paul Smeets[1], Robert Free[1], Lesley Brooker[3], Russell Chipman[2], and Derk Joester[1]*

Kaleidoscope Eyes: Microstructure and Optical Performance of Chiton Ocelli

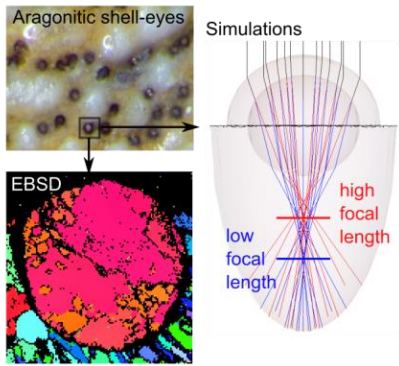

# Supplemental Information for Kaleidoscope Eyes: Microstructure and Optical Performance of Chiton Ocelli

Leanne Friedrich[1], Wai Sze Lam[2], Lyle Gordon[1], Paul Smeets[1], Robert Free[1], Lesley Brooker[3], Russell Chipman[2], and Derk Joester[1]*

[1] Materials Science and Engineering, Northwestern University, Evanston, IL 60208.
[2] College of Optical Sciences, University of Arizona, Tucson, AZ 85721.
[3] GeneCology Research Centre, University of the Sunshine Coast, Sippy Downs, QLD 4556, Australia.

* to whom correspondence should be addressed: d-joester@northwestern.edu

## S1.1 Microstructure

### S1.1.1 Geological crystal

Table S2. Geological crystals with at least 600 indexed points.

| ‹001› Kent distribution | EBSD map (inverse pole figure z coloring). |
|---|---|
| number of points 3373<br>mean direction (-0.28, -0.96, 0.02)<br>major axis (0.10, -0.01, 1.00)<br>minor axis (0.96, -0.28, -0.10)<br>κ 48960<br>β 9926<br>κ/β 4.933<br>StDev (θ) (˚) 0.3359<br>StDev (phi) (˚) 0.2184 | 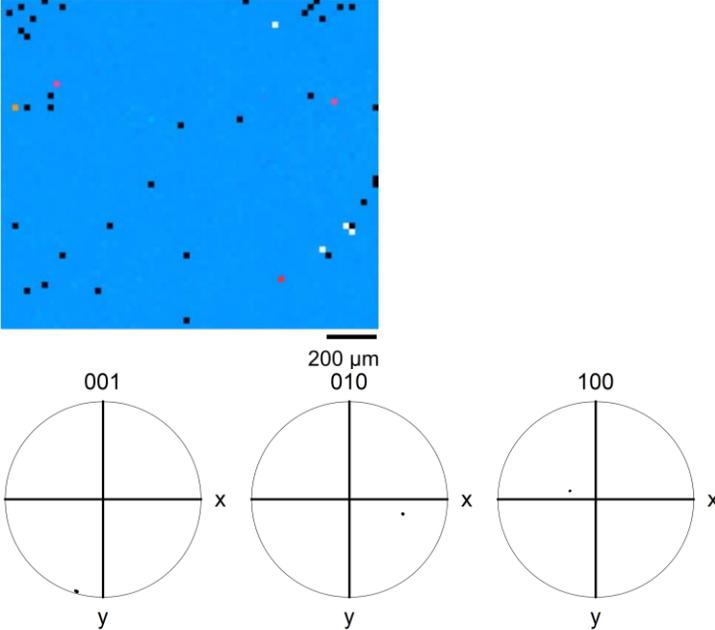 |
| number of points 2814<br>mean direction (0.18, 0.12, 0.98)<br>major axis (-0.96, -0.20, 0.20)<br>minor axis (0.21, -0.97, 0.08)<br>κ 8587<br>β 3141<br>κ/β 2.734<br>StDev (θ) (˚) 1.27<br>StDev (φ) (˚) 0.3357 | 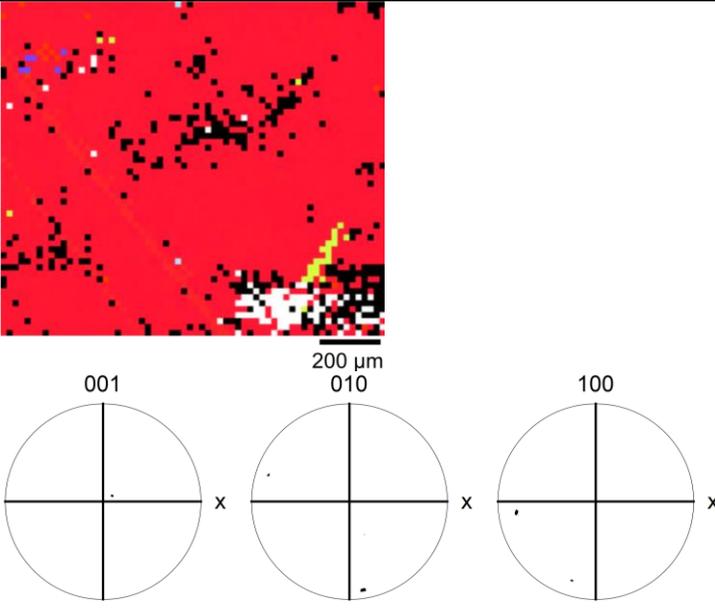 |
| number of points | 9191 |
| mean direction | (0.18, 0.13, 0.98) |
| major axis | (-0.97, -0.16, 0.20) |
| minor axis | (0.18, -0.98, 0.09) |
| κ | 4599 |
| β | 1349 |
| κ/β | 3.409 |
| StDev (θ) (°) | 1.527 |
| StDev (φ) (°) | 0.3448 |

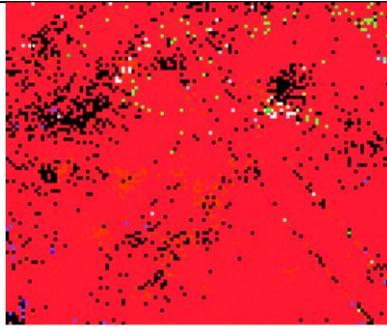
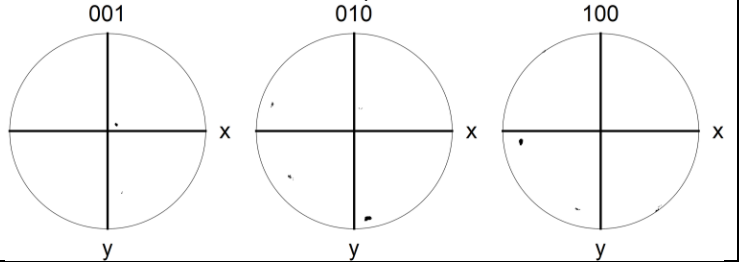

## S1.1.2 Oblique lenses

Table S3. Oblique lenses with at least 600 indexed points.

| ‹001› Kent distribution | | Optical micrograph, EBSD map (inverse pole figure z coloring). Images are not necessarily aligned. Approximate location of EBSD scan is shown. |
|---|---|---|
| number of points | 2217 | 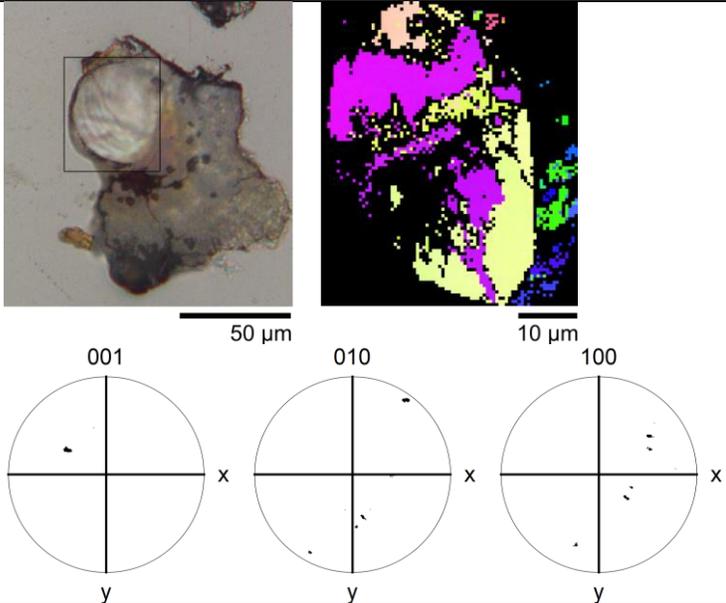 |
| mean direction | (-0.65, 0.41, 0.64) | |
| major axis | (-0.18, -0.90, 0.39) | |
| minor axis | (-0.74, -0.14, -0.66) | |
| κ | 1707 | |
| β | 359.6 | |
| κ/β | 4.747 | |
| StDev (θ) (°) | 1.841 | |
| StDev (φ) (°) | 1.16 | |
| number of points | 2187 | 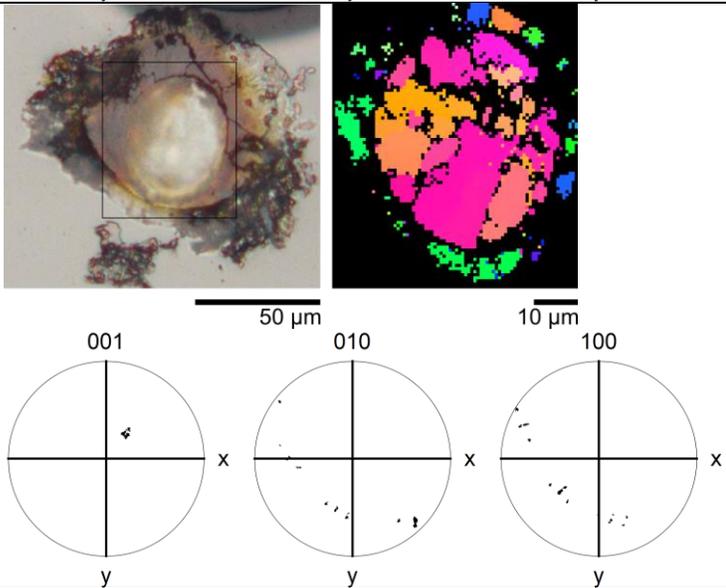 |
| mean direction | (0.35, 0.45, 0.82) | |
| major axis | (0.90, -0.43, -0.15) | |
| minor axis | (-0.28, -0.78, 0.55) | |
| κ | 635.2 | |
| β | 32.59 | |
| κ/β | 19.49 | |
| StDev (θ) (°) | 2.403 | |
| StDev (φ) (°) | 2.164 | |

## S1.1.3 Plan lenses
Table S4. Plan lenses with at least 600 indexed points.

| ‹001› Kent distribution | | Optical micrograph, EBSD map (inverse pole figure z coloring). Images are not necessarily aligned. Approximate location of EBSD scan is shown. |
|---|---|---|
| number of points | 11050 | 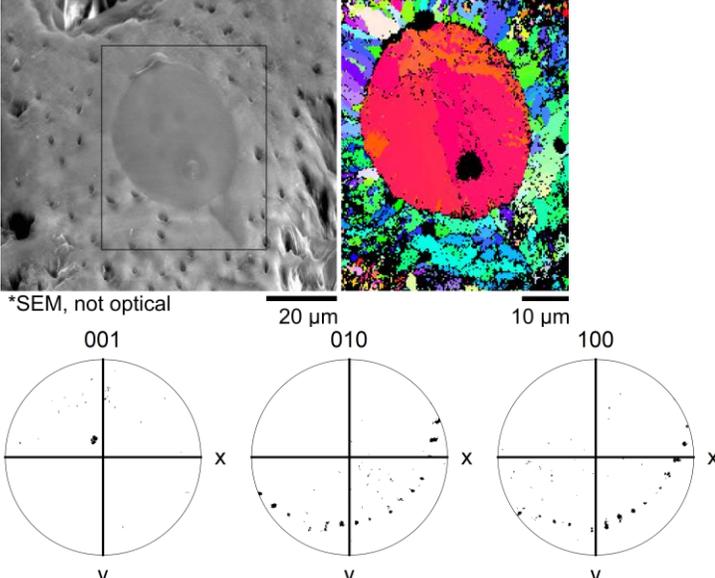 |
| mean direction | (-0.16, 0.34, 0.92) | |
| major axis | (-0.94, 0.24, -0.25) | |
| minor axis | (0.31, 0.91, -0.28) | |
| $\kappa$ | 230.7 | |
| $\beta$ | 35.99 | |
| $\kappa/\beta$ | 6.411 | |
| StDev ($\theta$) (°) | 4.851 | |
| StDev ($\varphi$) (°) | 3.227 | |
| number of points | 5148 | 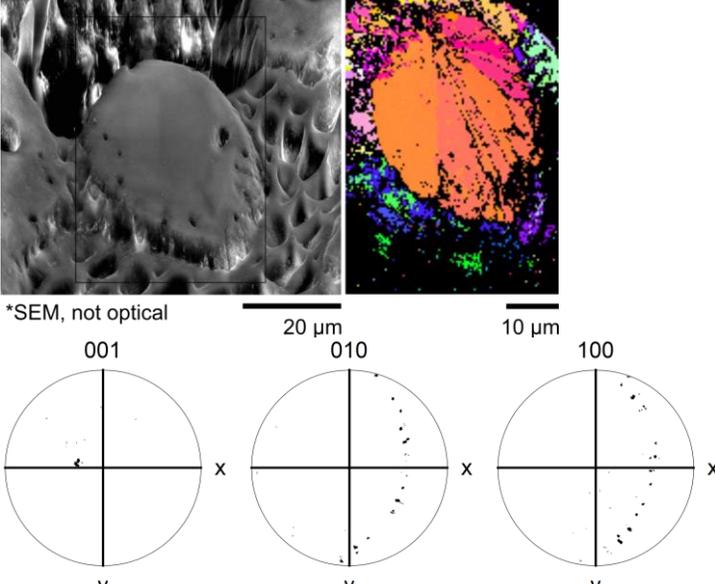 |
| mean direction | (-0.50, 0.07, 0.86) | |
| major axis | (0.83, -0.24, 0.50) | |
| minor axis | (0.24, 0.97, 0.06) | |
| $\kappa$ | 778.3 | |
| $\beta$ | 167.3 | |
| $\kappa/\beta$ | 4.652 | |
| StDev ($\theta$) (°) | 3.008 | |
| StDev ($\varphi$) (°) | 1.515 | |

| | |
|---|---|
| number of points | 5896 |
| mean direction | (0.70, 0.26, 0.66) |
| major axis | (0.40, 0.62, -0.67) |
| minor axis | (-0.59, 0.74, 0.33) |
| κ | 3087 |
| β | 1062 |
| κ/β | 2.907 |
| StDev (θ) (°) | 1.847 |
| StDev (φ) (°) | 0.793 |
| number of points | 7386 |
| mean direction | (0.26, -0.33, 0.91) |
| major axis | (0.02, -0.94, -0.35) |
| minor axis | (0.96, 0.11, -0.24) |
| κ | 3024 |
| β | 984 |
| κ/β | 3.073 |
| StDev (θ) (°) | 1.764 |
| StDev (φ) (°) | 0.8106 |

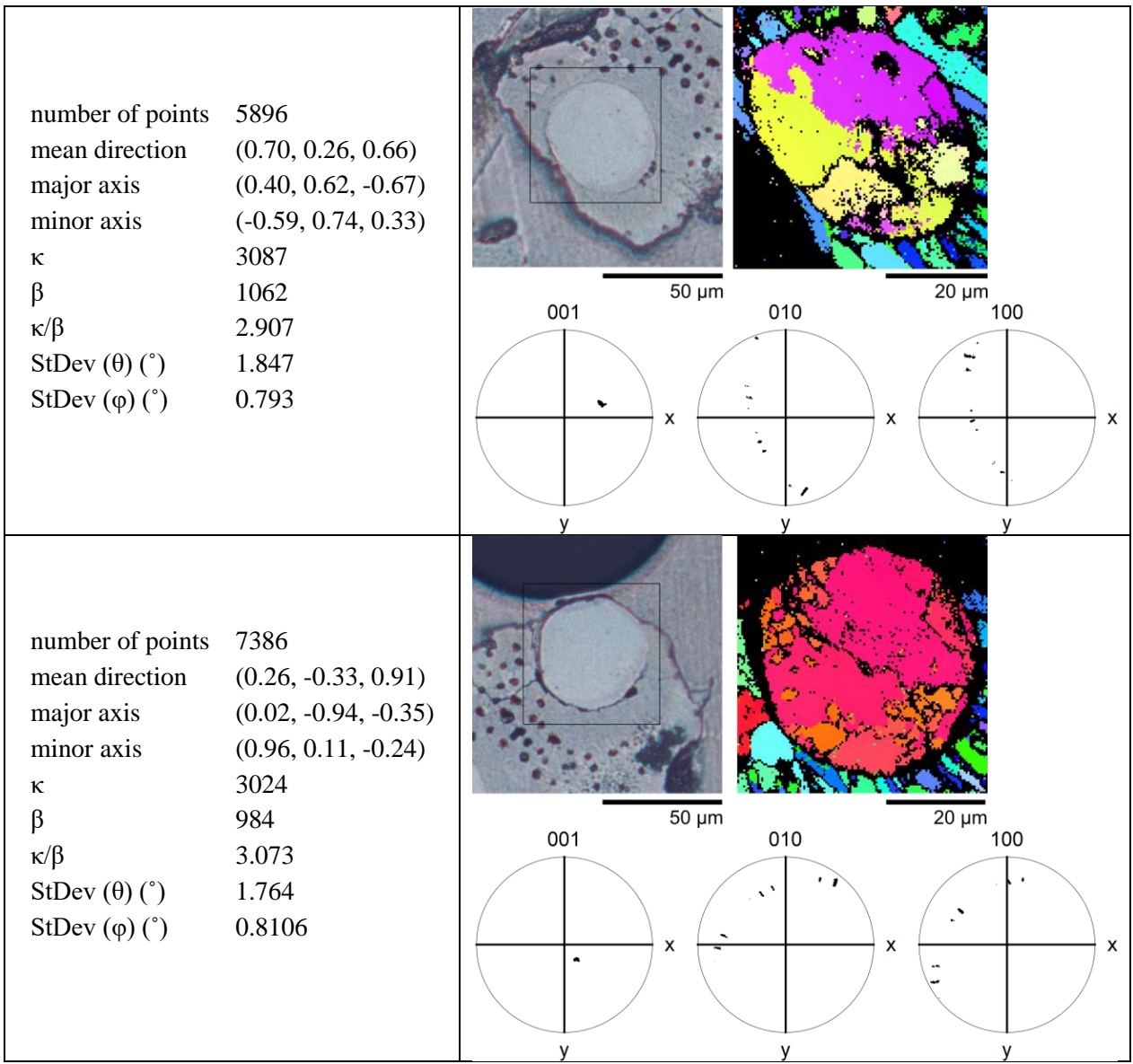

| | |
|---|---|
| number of points | 2993 |
| mean direction | (-0.70, -0.49, 0.41) |
| major axis | (0.61, -0.28, 0.74) |
| minor axis | (-0.26, 0.81, 0.52) |
| κ | 22.2 |
| β | 2.429 |
| κ/β | 9.137 |
| StDev (θ) (°) | 10.74 |
| StDev (φ) (°) | 5.085 |
| number of points | 616 |
| mean direction | (-0.38, 0.61, 0.70) |
| major axis | (-0.59, -0.74, 0.33) |
| minor axis | (-0.72, 0.29, -0.64) |
| κ | 4953 |
| β | 1827 |
| κ/β | 2.711 |
| StDev (θ) (°) | 1.591 |
| StDev (φ) (°) | 0.6177 |

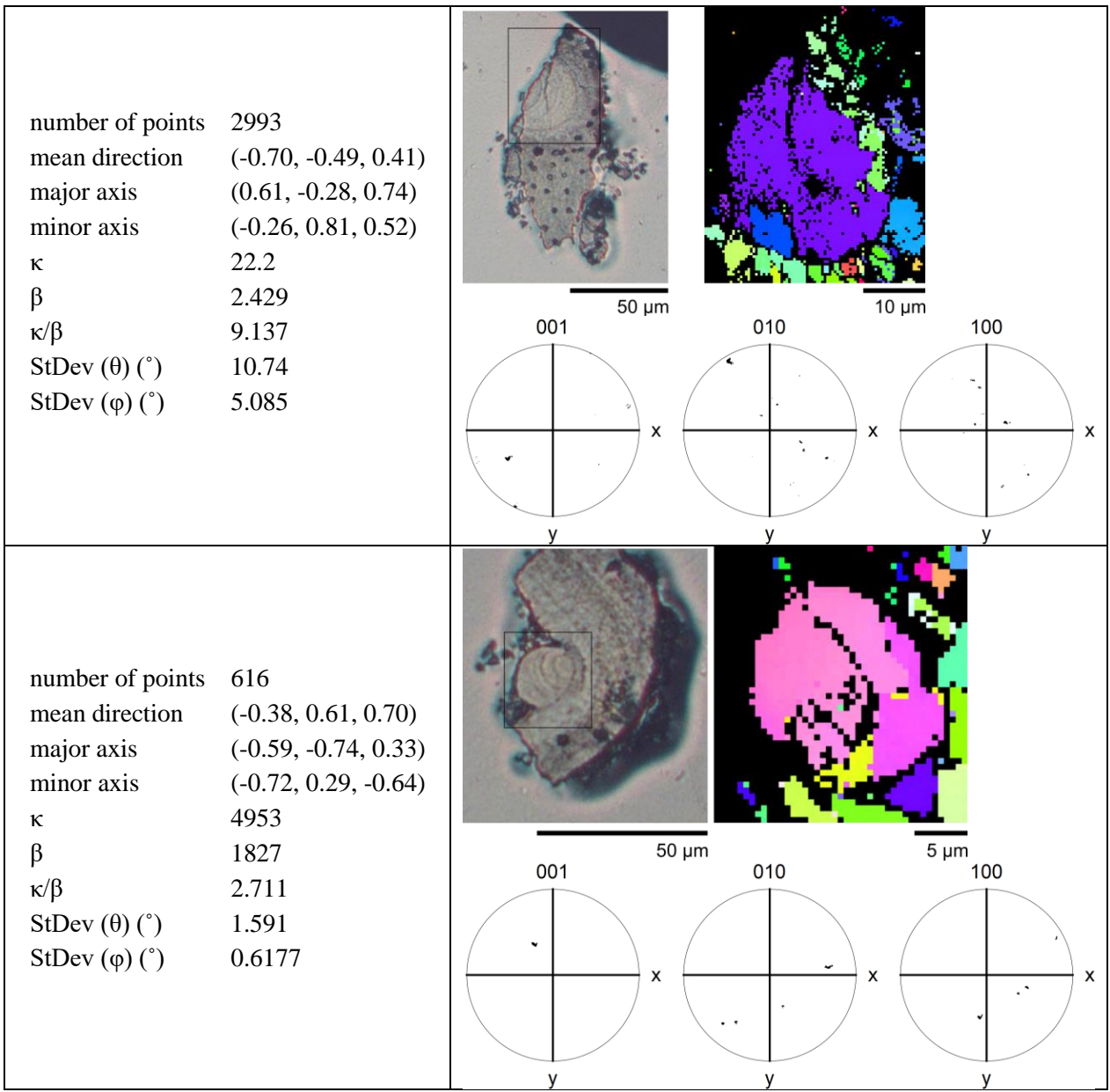

| | |
|---|---|
| number of points | 6381 |
| mean direction | (0.41, 0.89, 0.21) |
| major axis | (-0.55, 0.43, -0.71) |
| minor axis | (0.72, -0.18, -0.67) |
| κ | 3600 |
| β | 779.8 |
| κ/β | 4.616 |
| StDev (θ) (°) | 1.279 |
| StDev (φ) (°) | 0.7938 |
| number of points | 2266 |
| mean direction | (0.56, -0.78, 0.28) |
| major axis | (-0.12, 0.26, 0.96) |
| minor axis | (0.82, 0.57, -0.05) |
| κ | 3073 |
| β | 354.4 |
| κ/β | 8.67 |
| StDev (θ) (°) | 1.179 |
| StDev (φ) (°) | 0.9318 |
| number of points | 3842 |
| mean direction | (-0.52, -0.18, 0.83) |
| major axis | (0.08, -0.98, -0.16) |
| minor axis | (-0.85, 0.02, -0.53) |
| κ | 599.2 |
| β | 111.1 |
| κ/β | 5.392 |
| StDev (θ) (°) | 3.01 |
| StDev (φ) (°) | 2.034 |

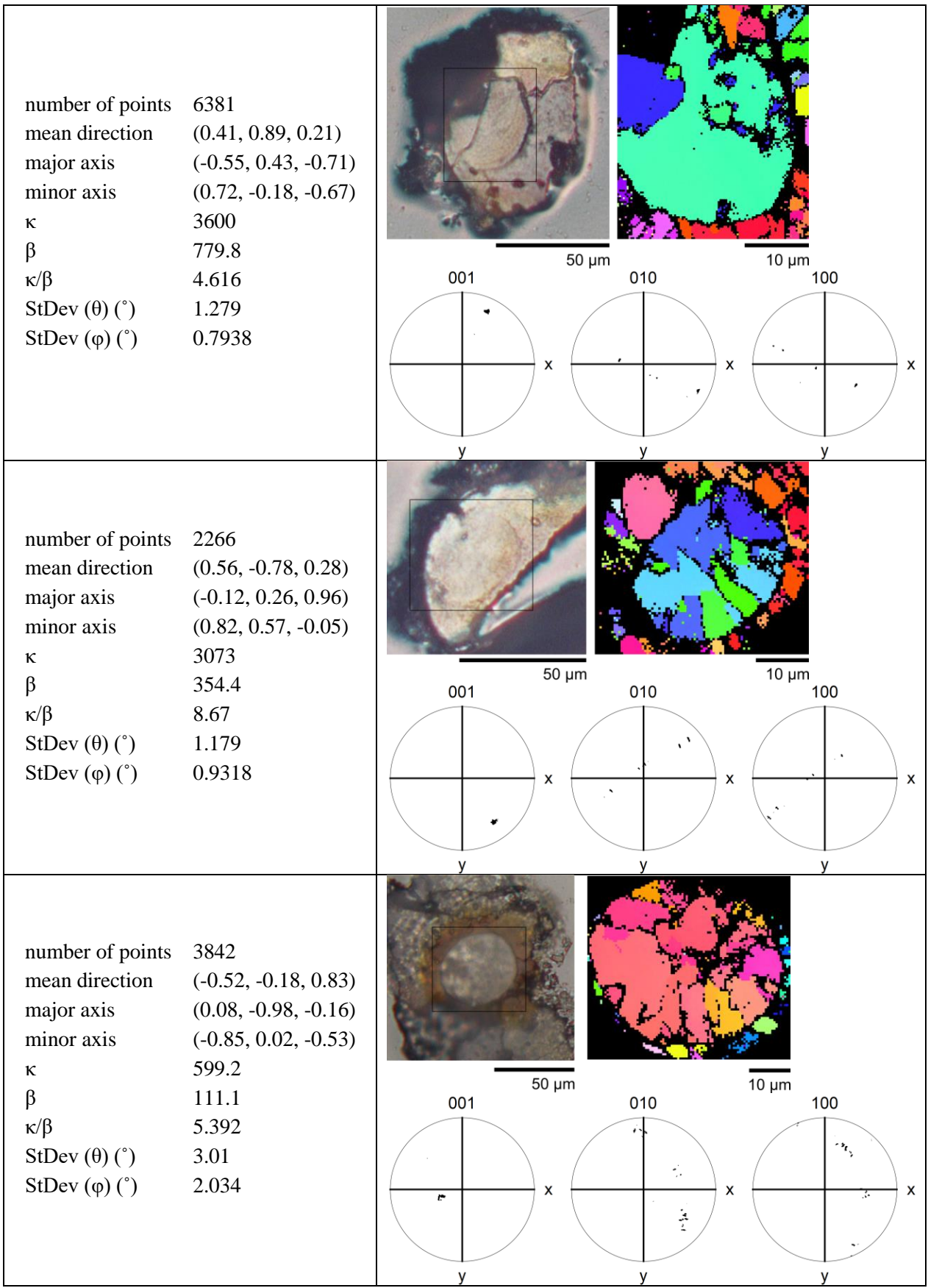

| | |
|---|---|
| number of points | 4437 |
| mean direction | (-0.23, -0.61, 0.75) |
| major axis | (0.83, -0.53, -0.18) |
| minor axis | (0.51, 0.59, 0.63) |
| κ | 249 |
| β | 35.1 |
| κ/β | 7.094 |
| StDev (θ) (°) | 4.794 |
| StDev (φ) (°) | 2.737 |
| number of points | 2070 |
| mean direction | (-0.11, 0.34, 0.93) |
| major axis | (0.72, 0.67, -0.16) |
| minor axis | (0.68, -0.66, 0.32) |
| κ | 4020 |
| β | 366.9 |
| κ/β | 10.96 |
| StDev (θ) (°) | 0.9998 |
| StDev (φ) (°) | 0.8312 |

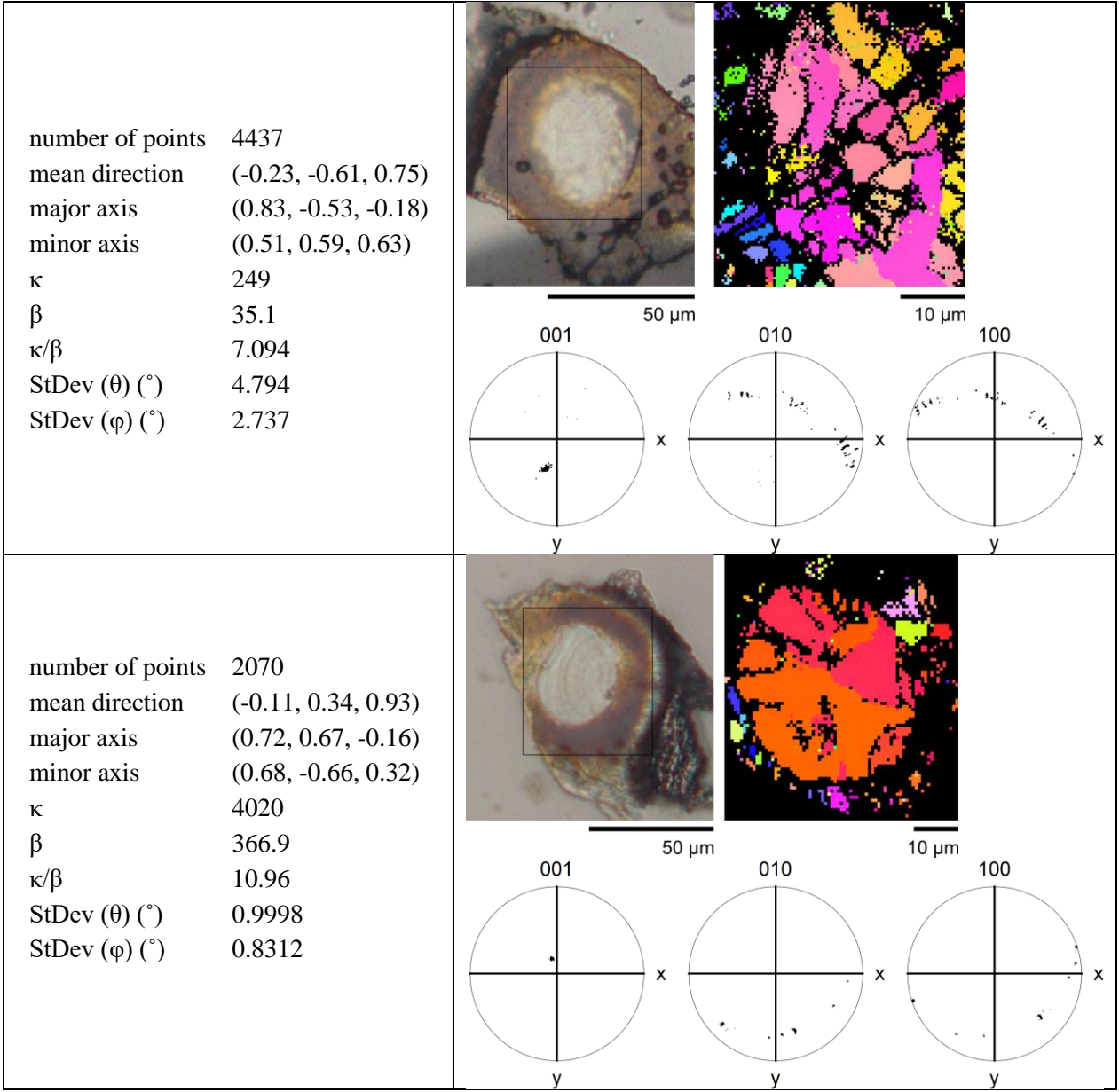

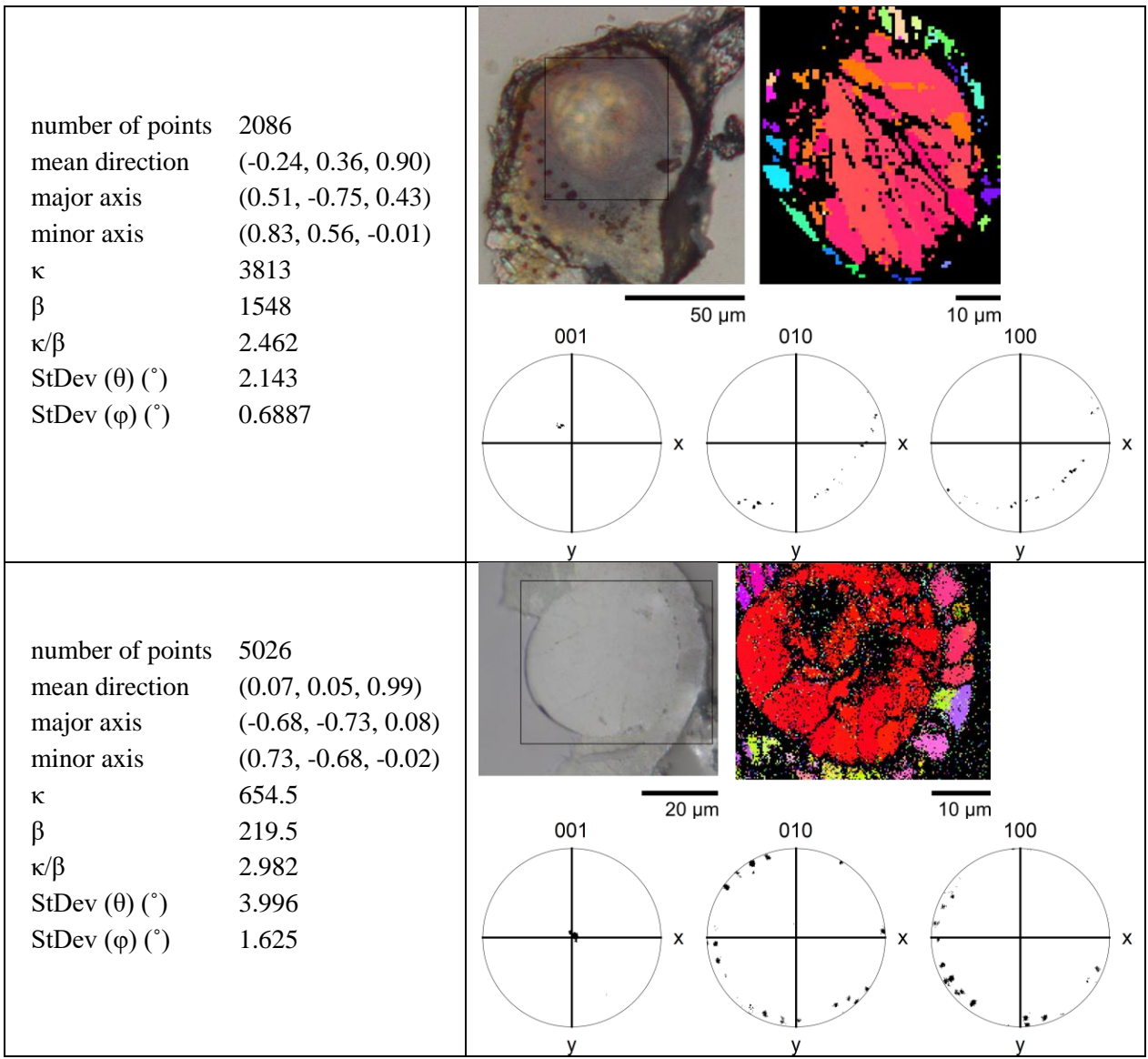

## S1.1.4 Cross-sections

Table S5. Cross-sections with at least 600 indexed points.

| ‹001› Kent distribution | Optical micrograph, EBSD map (inverse pole figure z coloring). Images are not necessarily aligned. Approximate location of EBSD scan is shown. |
|---|---|
| This sample was indexed using an incorrect aragonite crystal data file and was not included in crystallographic analyses. | 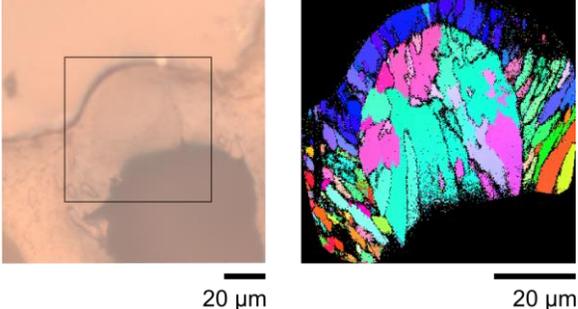 |
| number of points   7252<br>mean direction   (-0.14, 0.63, 0.76)<br>major axis   (-0.97, 0.08, -0.25)<br>minor axis   (0.22, 0.77, -0.60)<br>κ   212.9<br>β   35.56<br>κ/β   5.986<br>StDev (θ) (˚)   5.405<br>StDev (φ) (˚)   1.788 | 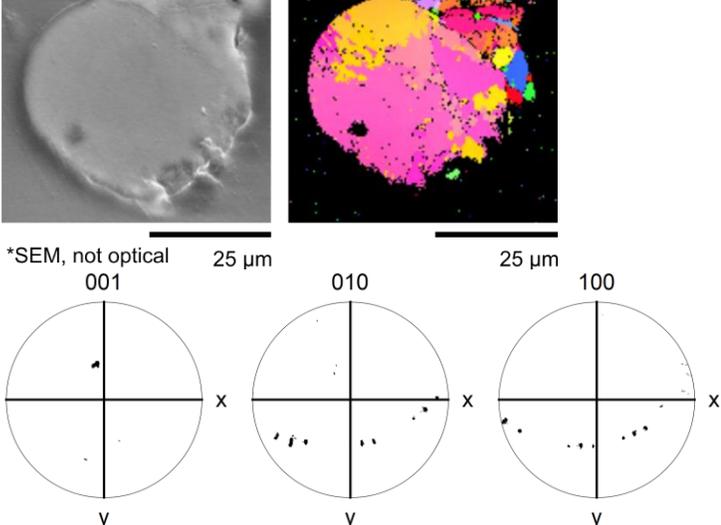 |
| This sample was indexed using an incorrect aragonite crystal data file and was not included in crystallographic analyses. | 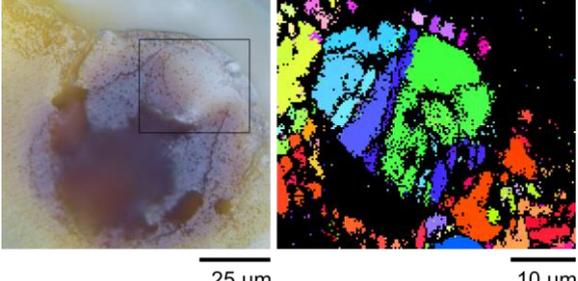 |
| number of points | 647 |
| mean direction | (0.76, -0.37, 0.39) |
| major axis | (-0.06, 0.61, 0.79) |
| minor axis | (0.55, 0.68, -0.49) |
| $\kappa$ | 14.02 |
| $\beta$ | 0.8973 |
| $\kappa/\beta$ | 15.62 |
| StDev $(\theta)$ (°) | 18.24 |
| StDev $(\varphi)$ (°) | 9.889 |
| number of points | 4492 |
| mean direction | (-0.41, 0.88, 0.26) |
| major axis | (-0.13, 0.22, -0.97) |
| minor axis | (-0.90, -0.43, 0.02) |
| $\kappa$ | 2381 |
| $\beta$ | 642 |
| $\kappa/\beta$ | 3.709 |
| StDev $(\theta)$ (°) | 1.731 |
| StDev $(\varphi)$ (°) | 0.9459 |
| number of points | 694 |
| mean direction | (-0.18, 0.98, 0.13) |
| major axis | (-0.60, 0.00, -0.80) |
| minor axis | (-0.78, -0.22, 0.59) |
| $\kappa$ | 3550 |
| $\beta$ | 1030 |
| $\kappa/\beta$ | 3.446 |
| StDev $(\theta)$ (°) | 1.486 |
| StDev $(\varphi)$ (°) | 0.7652 |

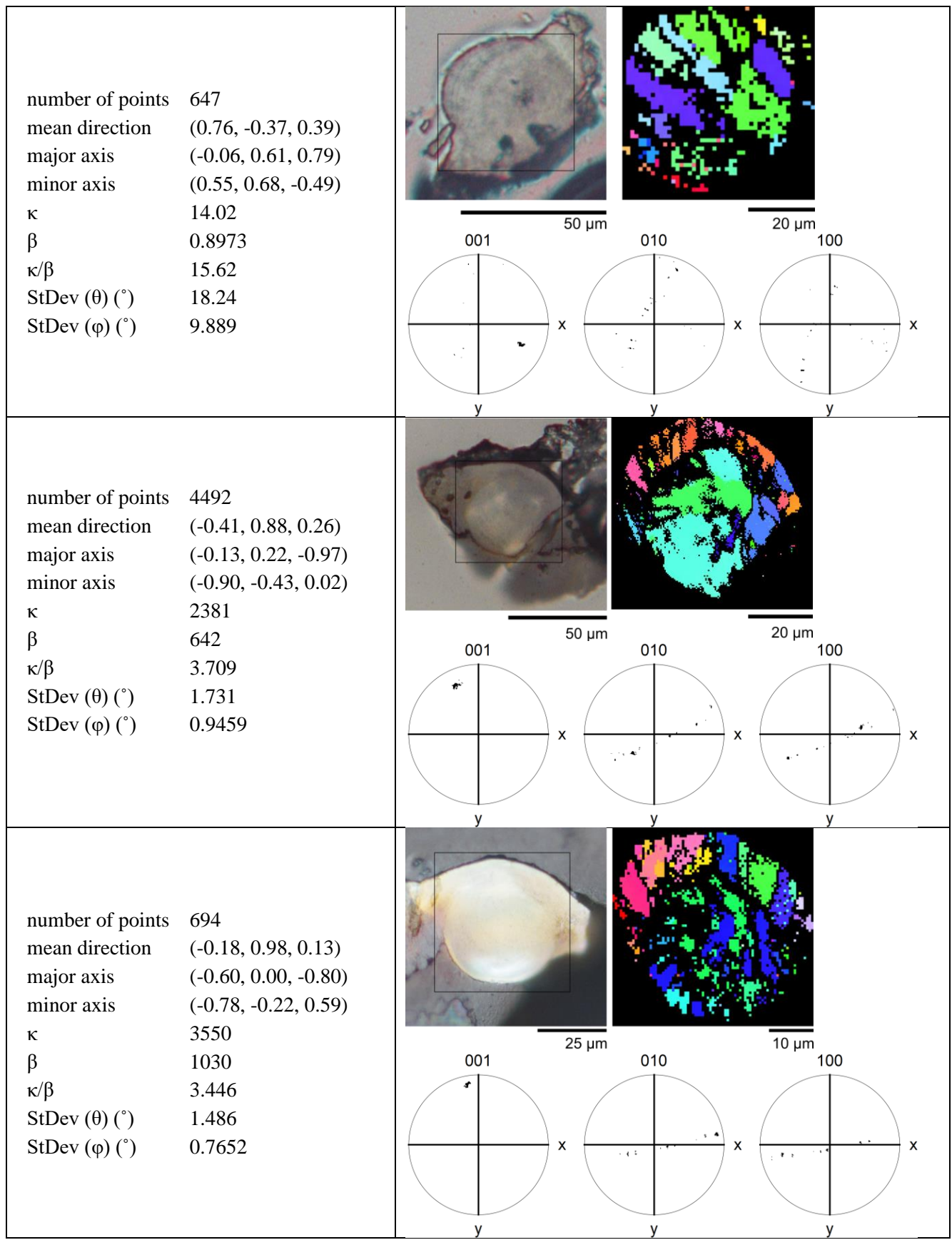

| number of points | 1973 |
| --- | --- |
| mean direction | (-0.74, -0.13, 0.65) |
| major axis | (-0.58, -0.37, -0.73) |
| minor axis | (0.34, -0.92, 0.21) |
| κ | 2822 |
| β | 1252 |
| κ/β | 2.254 |
| StDev (θ) (°) | 3.216 |
| StDev (φ) (°) | 0.7791 |

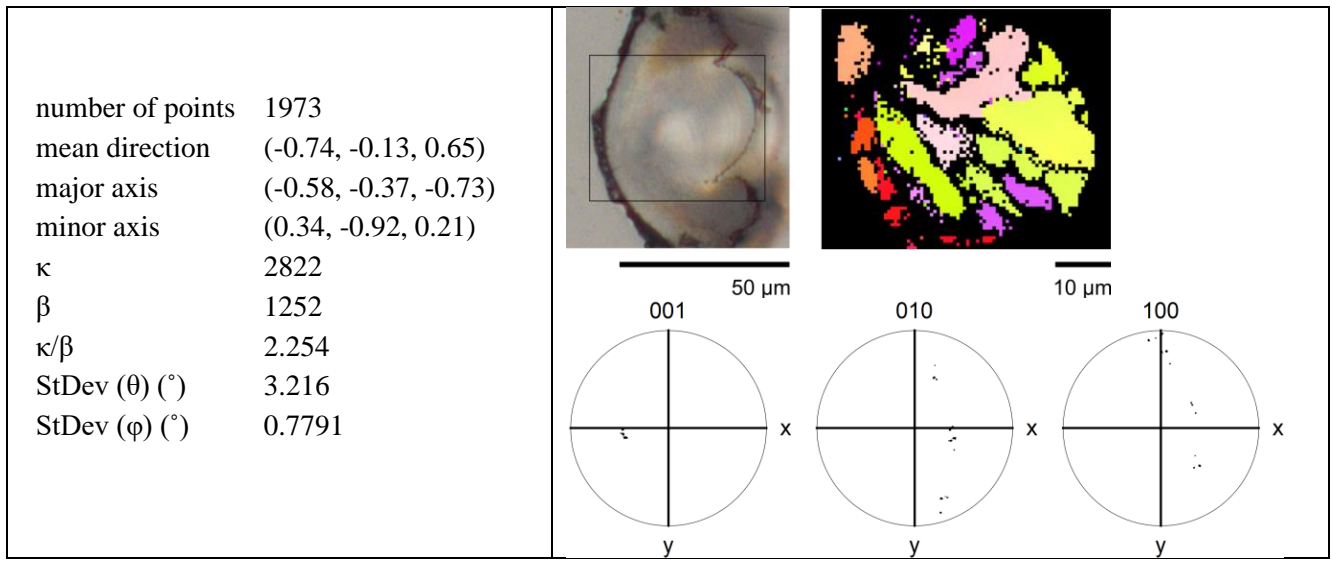

### S1.1.5 Shell

Table S6. Shells with at least 600 indexed points.

| ‹001› Kent distribution | EBSD map (inverse pole figure z coloring). |
| --- | --- |
| number of points 5980<br>mean direction (0.26, 0.10, 0.86)<br>major axis (0.60, -0.80, -0.10)<br>minor axis (-0.75, -0.60, 0.30)<br>κ 10.24<br>β 0.1981<br>κ/β 51.67<br>StDev (θ) (°) 20<br>StDev (φ) (°) 17.18 | 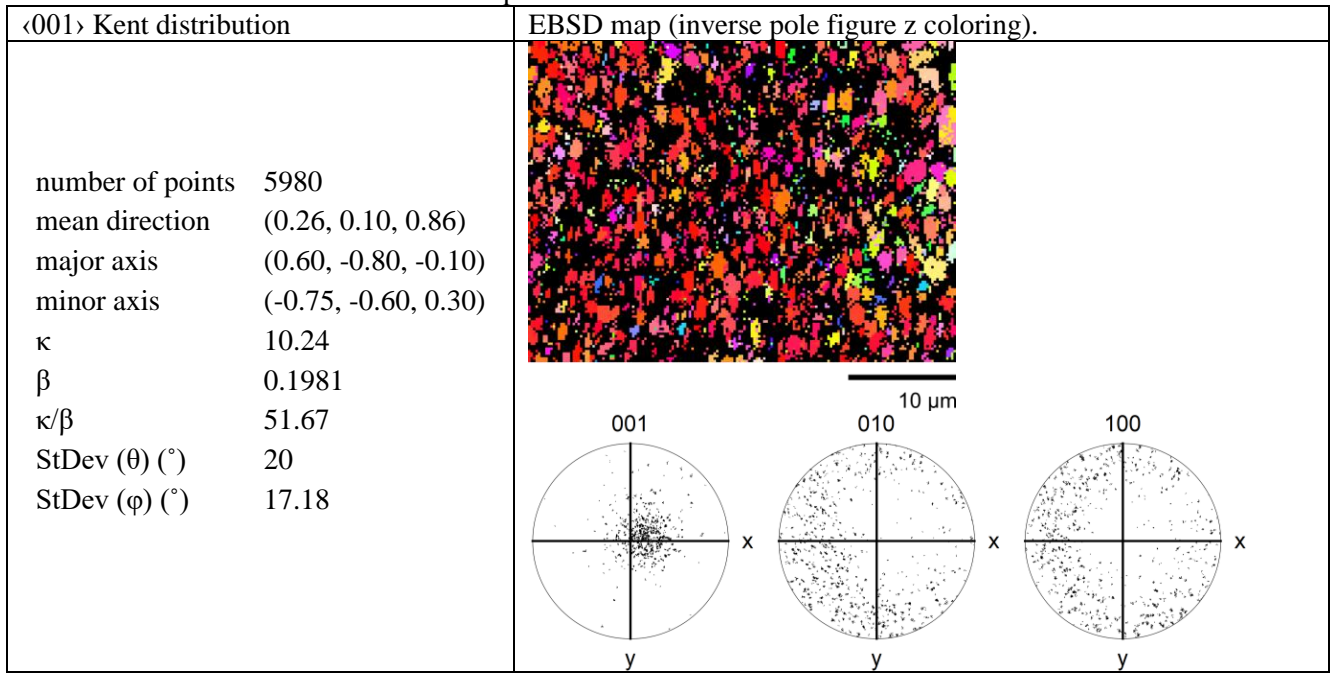 |

| Table S7. Standard deviations of single grain ⟨001⟩ orientations within samples. | | | |
|---|---|---|---|
| Type | # pixels | Long STDEV (°) | Short STDEV (°) |
| Geological | 3546 | 0.352 | 0.272 |
| Geological | 3237 | 0.326 | 0.217 |
| Geological | 211 | 0.369 | 0.274 |
| Geological | 1484 | 0.361 | 0.260 |
| Lens | 665 | 1.066 | 0.435 |
| Lens | 387 | 0.641 | 0.304 |
| Lens | 728 | 1.767 | 0.283 |
| Lens | 1038 | 0.701 | 0.304 |
| Lens | 523 | 0.487 | 0.261 |
| Lens | 3641 | 0.912 | 0.406 |
| Lens | 1780 | 0.479 | 0.386 |
| Lens | 2484 | 1.679 | 0.455 |
| Lens | 4766 | 1.884 | 0.713 |
| Lens | 2281 | 0.603 | 0.424 |
| Lens | 232 | 0.999 | 0.699 |
| Lens | 5017 | 1.149 | 0.630 |
| Lens | 532 | 0.973 | 0.339 |
| Lens | 1044 | 0.755 | 0.568 |
| Lens | 3026 | 1.420 | 0.505 |
| Lens | 130 | 0.813 | 0.642 |
| Lens | 1949 | 1.069 | 0.498 |
| Lens | 503 | 0.684 | 0.231 |
| Shell | 531 | 0.621 | 0.334 |
| Shell | 383 | 0.379 | 0.256 |
| Shell | 548 | 0.402 | 0.337 |
| Shell | 319 | 0.350 | 0.259 |
| Shell | 229 | 0.858 | 0.445 |
| Shell | 159 | 0.305 | 0.227 |
| Shell | 438 | 0.736 | 0.306 |
| Shell | 195 | 0.372 | 0.260 |
| Shell | 161 | 0.306 | 0.242 |
| Shell | 124 | 0.550 | 0.380 |
| Shell | 129 | 0.512 | 0.358 |

| Table S8. ANOVA tables of long standard deviation as a function of region (lens, valve, geological). Because the p-value is below 0.05, these data indicate a significant difference in single-grain overall ⟨001⟩ axis orientation distribution width between lenses, valves, and geological crystals. | | | | | |
|---|---|---|---|---|---|
| Long | DF | SumOfSq | MeanSq | FRatio | PValue |
| Model | 2 | 2.54172 | 1.27086 | 10.9187 | 0.000274 |
| Error | 30 | 3.4918 | 0.116393 | | |
| Total | 32 | 6.03351 | | | |

| Table S9. ANOVA table of short standard deviation as a function of region (lens, valve, geological). Because the p-value is below 0.05, these data indicate a significant difference in single-grain ⟨001⟩ axis orientation distribution width between lenses, valves, and geological crystals. | | | | | |
|---|---|---|---|---|---|
| Short | DF | SumOfSq | MeanSq | FRatio | PValue |
| Model | 2 | 0.202699 | 0.10135 | 6.8751 | 0.003484 |
| Error | 30 | 0.442247 | 0.014742 | | |
| Total | 32 | 0.644946 | | | |

Table S10. Average sizes of multi-grain ⟨001⟩ axis orientation Kent distributions across all samples. Shells have the widest distributions, while geological crystals have the narrowest distributions. *N* indicates the number of EBSD maps included in the measurement.

|  | geological (*N*=4) | lens (*N*=18) | shell (*N*=11) |
| --- | --- | --- | --- |
| $\sigma_{short}$ [°] | 0.31 ± 0.06 | 1.96 ± 2.14 | 14.13 ± 5.76 |
| $\sigma_{long}$ [°] | 0.89 ± 0.60 | 4.05 ± 4.16 | 27.84 ± 8.92 |

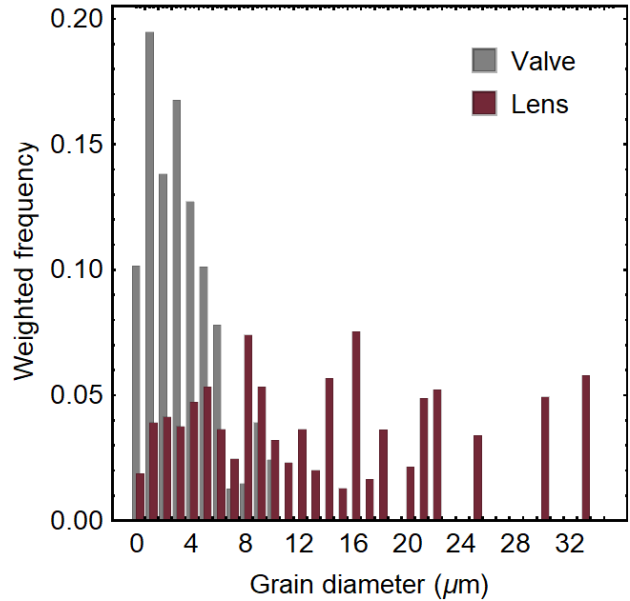

Figure S1. Frequency of grain sizes in the lens and valve, weighted by grain area.

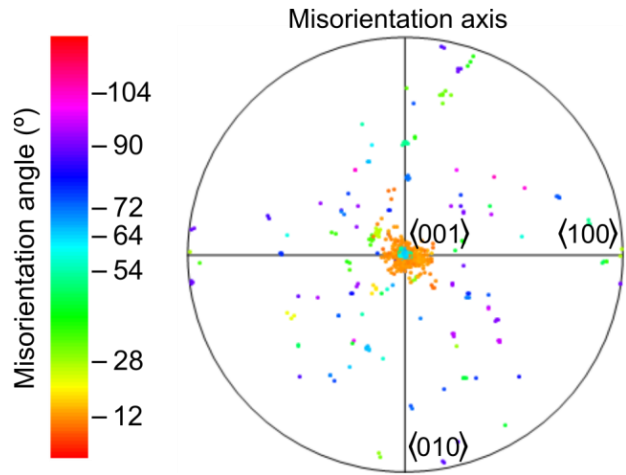

Figure S2. Pole figure of misorientation axes as a function of misorientation angle for the lens grain boundaries in Figure 3B.

## S1.2 The cornea

The cornea can only be distinguished in cross sections and often fractures during extraction from the valve, so our analysis is based on a small number of well-preserved examples (*N* = 5). In EBSD orientation maps, the cornea appears as a single layer of roughly columnar grains. The width of the cornea decreases from the edge, where it fuses with the valve, to the center (7 ± 2 µm). The ⟨001⟩ is somewhat aligned in the cornea, likely due to the small number of available grains (Figure S4).

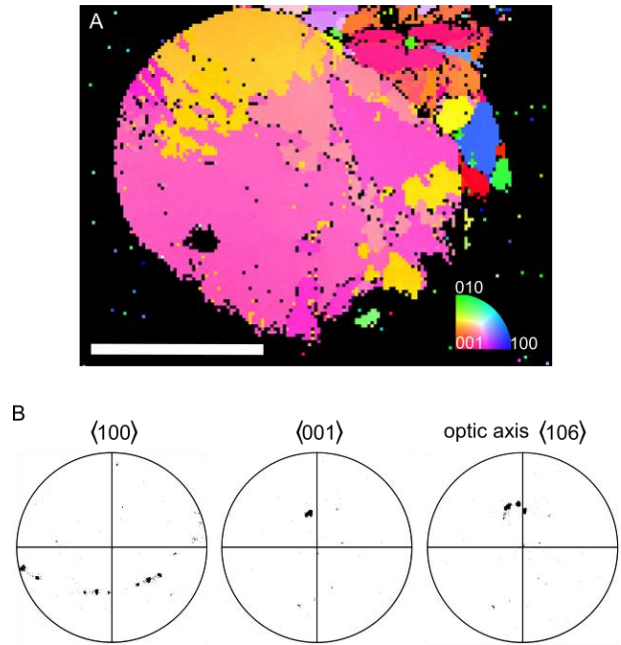

Figure S3. **A)** Cross section, with missing cornea at the bottom right. Scale bar represents 20 µm. **B)** Equal area projection pole figures for lens in A.

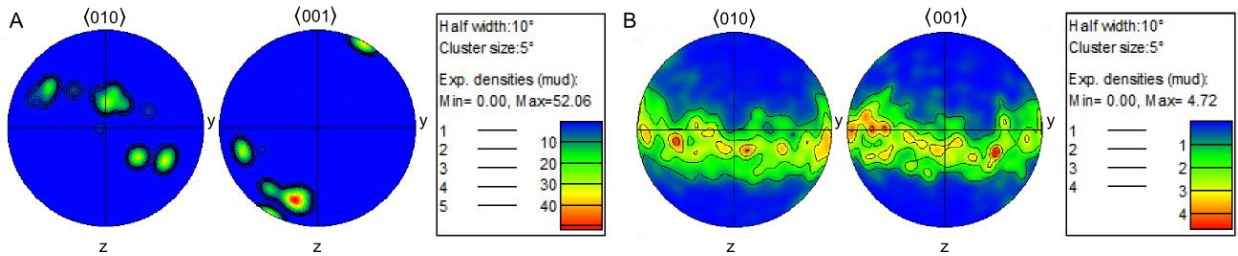

Figure S4. Examples of contoured pole figures for aragonite grains in (A) cornea and (B) valve. Few samples had complete corneas (N=5), and within those corneas, only a few grains are present. This contributes to the apparent texture in A.

## S1.3 Etching

Etched lenses exhibit a series of concentric grooves (Figure S5C,D). The innermost groove surrounds a region of the lens with a diameter of 5-15 µm. This core structure appears to have a finely granular surface structure in etched sections (Figure S5G). In EBSD maps, large portions of interfaces between grains in the core often cannot be indexed (**Figure 5**, Figure S5B,C). This may indicate that the core consists of grains which are finer than the resolution of the scan. In the periphery, the appearance of the etched surfaces depends on the density of grooves. Where few are present, deep, faceted etch pits that resemble trenches are found in an otherwise smooth surface (Figure S5E,H,K). Trenches have a width of 100-300 nm, a length of 1-2 µm and are oriented approximately radially. In contrast, regions with a high density of grooves exhibit higher surface roughness and snaking trenches that are shallower and narrower than the previous kind (Figure S5F,I,L).

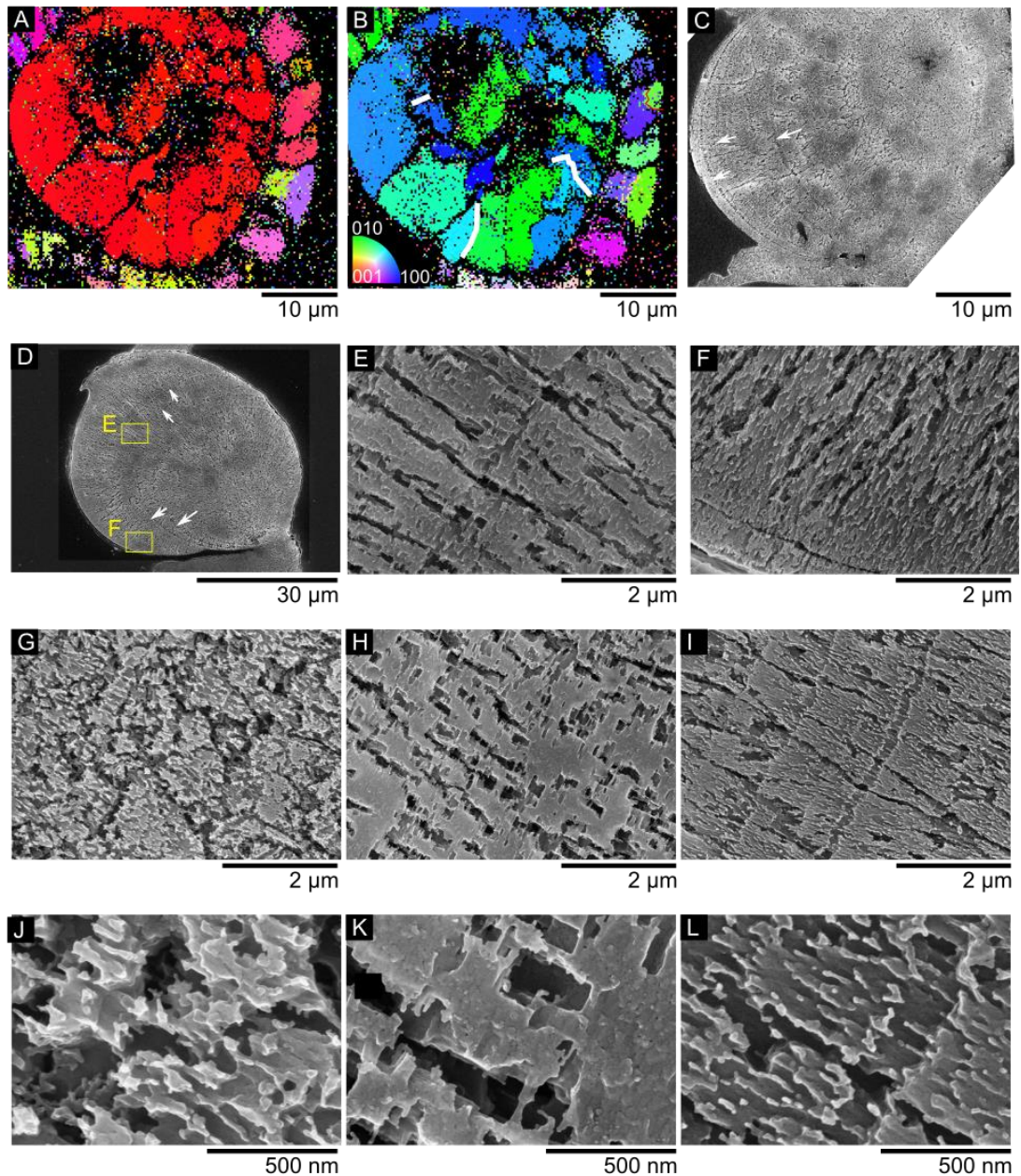

Figure S5. **A-C)** EBSD and etching of lens which was polished, etched, imaged, and then re-polished and imaged using EBSD. **A)** Inverse pole figure z-axis scan of polished lens. **B)** Inverse pole figure x-axis scan of polished lens. White lines indicate twin interfaces. **C)** SEM micrograph of etched lens surrounded partially by shell. Scale bars 10 µm. Arrows indicate etch bands. **D-L)** Etched lens sections. **D)** Lens. **E)** Deep etch pits in region with low band density. **F)** Shallow, curved etch pits in region with high band density. **G,J)** Core of etch bands. **H,K)** Outer region, with low etch band density and deep, faceted etch pits. **I,L)** Outer region with high etch band density and shallow, curved etch pits.

## Simulations

Rays entering the cornea from air are refracted more strongly than rays entering from water because air has a lower refractive index than water. As a result, lenses in air have a shorter focal length than lenses in water. However, unlike in isotropic lenses, in anisotropic lenses the focal length ratio between air and water is not constant. This ratio varies between points at equivalent radii (Figure S6B) and for varying ⟨001⟩ orientations (Figure S6A). Thus, because of this nonlinearity which comes from the crystallography of the lens, it is impossible to predict how the focal length and aberrations will change between air and water without knowing the orientation of the ⟨001⟩ axis.

Because the entire cell cavity is full of sensory cells, it is well-suited to detect focused light whether the chiton is on land or underwater.

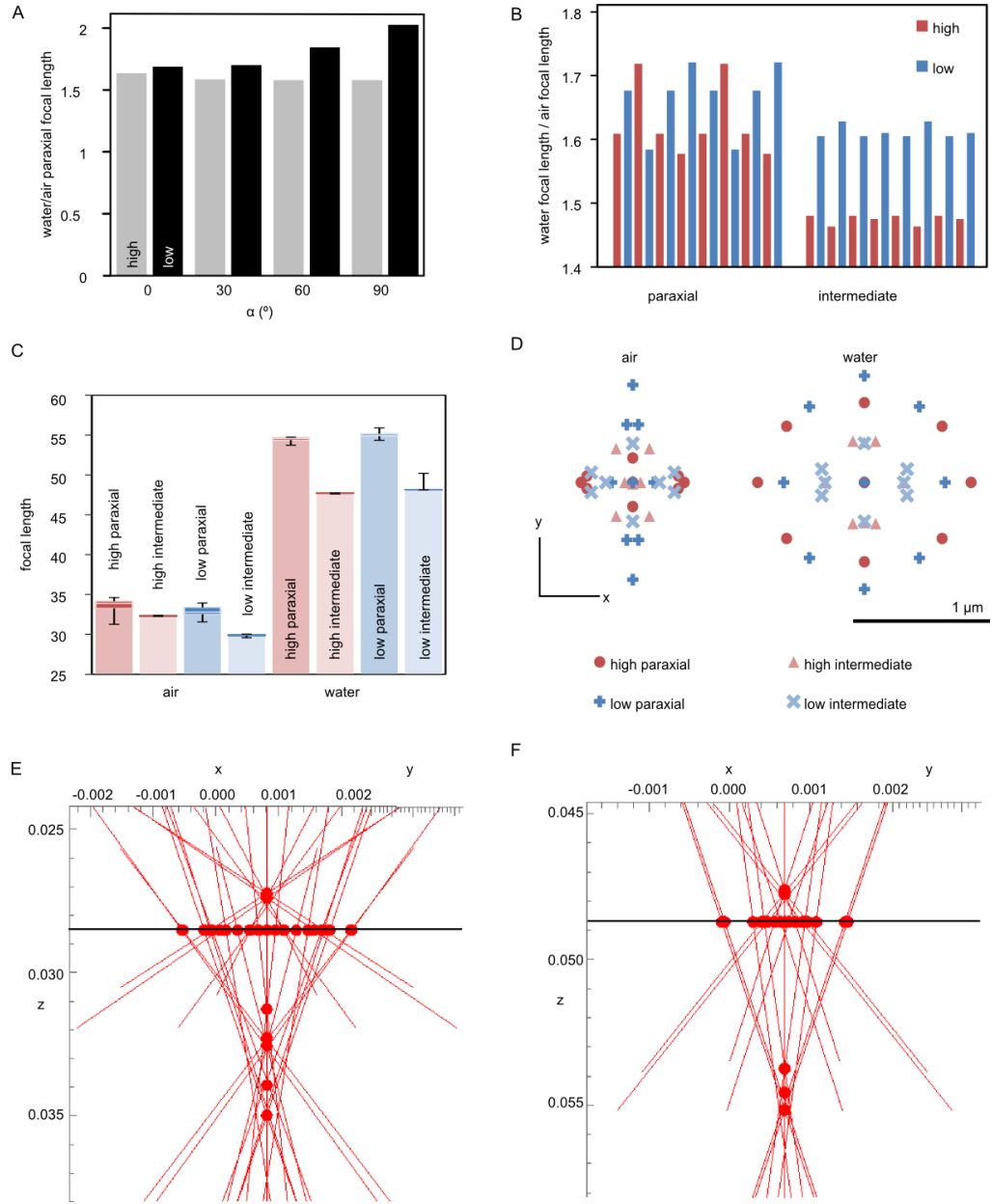

Figure S6. **A)** Paraxial focal length in air divided by the paraxial focal length in water. **B)** Ratios of individual ray focal lengths in water to corresponding focal lengths in air for α=0º. **C)** Focal lengths of α=0º lenses in air and water. **D)** Intersections between rays and the plane of least confusion. Both sets are centered on the lens optical axis. **E, F)** High ray focal regions, viewed 70º away from the ⟨010⟩ projection so all rays can be seen. Black line indicates plane of least confusion. Dots on the black line indicate intersections with the plane of least confusion. All other dots are focal points. Lens in (E) air and (F) water. Scale is in mm.

Introducing a horizontal twin grain boundary at the center of the lens changes the paraxial focal lengths of the lens by no more than 0.16 µm, and it reduces longitudinal aberration (Figure S7B). In contrast, the non-twin grain boundary has a larger low focal length and low ray aberrations, compared to the single-crystal lens (Figure S7B, C). Because the non-twin interface changes the low focal length of the lens while the twin interface does not create large changes, maintaining a high number of twin grain boundaries and low number of non-twin grain boundaries is important.

Simulations show that if a lens has an overall ⟨001⟩ orientation between 0º and 90º, the low rays can experience horizontal displacement, which depends on the ⟨100⟩ orientation of the lattice. If the lens is split into two twinned grains, the grains would have ⟨100⟩ orientations that vary by 64º, so the horizontal displacement directions of the two grains would also vary by 64º. In effect, this could create four focal points (two per grain), which would produce two non-displaced images from high rays and two horizontally displaced images from low rays (Figure S8). Introducing a vertical twin grain boundary to a lens where $\alpha > 0$ produces multiple focal points, where each grain of a different ⟨100⟩ orientation adds a new low ray focal point (Figure S8). Thus, the multiple images seen in Figure S9 could have come from twinned and non-twinned grain boundaries.

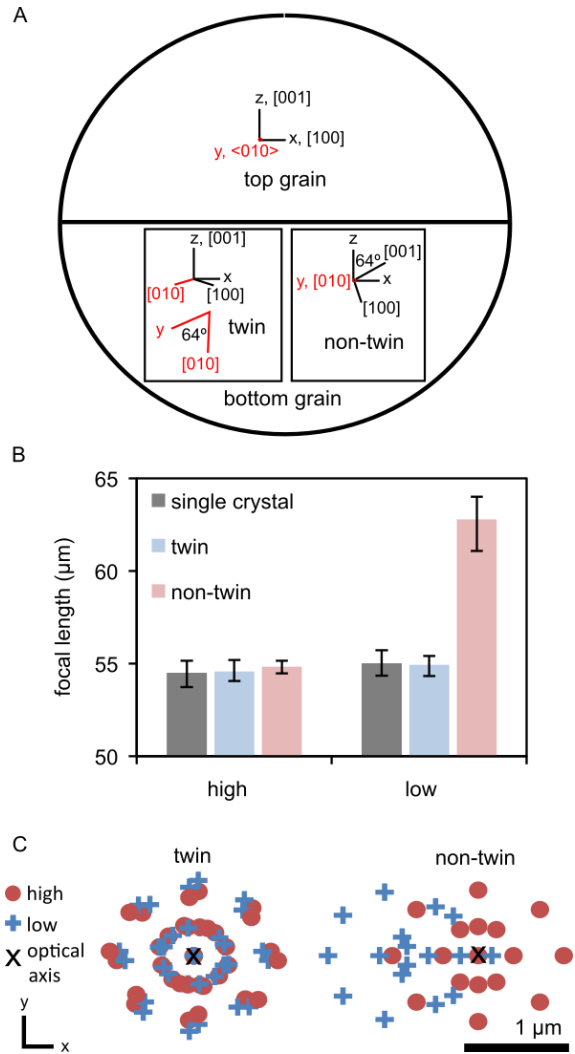

Figure S7. **A)** Diagram of twinned lens. **B)** Paraxial focal lengths of lenses. Error bars represent maximum and minimum focal lengths. **C)** Ellipse of least confusion for twin and non-twin lenses. "High" and "Low" indicate the eigenstate in the first grain. Rays that change eigenstate between grains are not included for the non-twin lens because they are not transmitted.

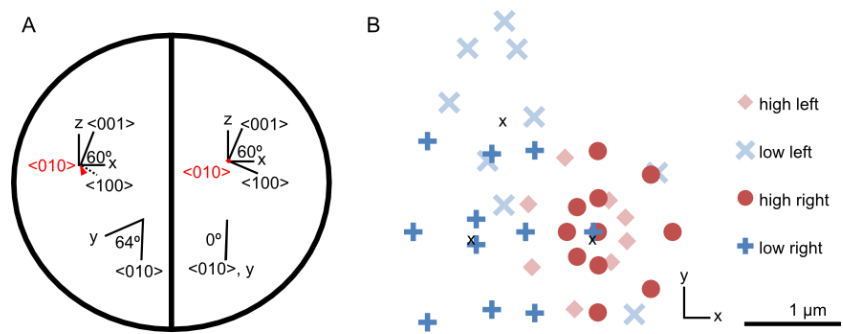

Figure S8. **A)** Lens (α = 30°) containing two grains, twinned along the y-z plane. **B)** Intersections between rays and the ellipse of least confusion. Right rays traveled through the right half of the lens, and left rays traveled through the left half of the lens. Three horizontally distinct focal points (high rays, low left rays, and low right rays, marked as black X) are visible.

## S1.4 Image Transmission

Dozens of ocelli lenses were extracted from shells using a razor blade, placed on a glass slide in air, and covered with a cover slip. The edges of the cover slip were sealed to the glass slide using nail polish. For comparison, amorphous calcium carbonate (ACC) microlens arrays were prepared according to Lee (2012).[50] Briefly, 1 g of CaOH was dissolved in 100 mL water in a sealed jar. After three days, 23.4 µL of an aqueous solution of Polysorbate 20 (0.22 M) was added to the solution, which was then stirred vigorously. The solution was apportioned into 40 mm petri dishes. After 1 hour, microlens films formed at the surface of the solution. The films were skimmed off the tops of the dishes onto a cover slip, and residual water was wicked from the side of the cover slip using lint-free paper. Cover slips were dried in air and then placed film-side-down onto glass slides.

A mask was created by painting a glass slide with opaque nail polish and carving an "N" shape into the dried paint (Figure S9A). The mask was placed directly on top of the light source of a Leica DM4000 upright microscope (Figure S10). Using Köhler alignment, the image of the mask was focused below the ocellus lens such that the entire microscope field of view was evenly illuminated.

To evaluate the performance of the lens, we transmitted images through isolated lenses. Optical performance varies between lenses. In Figure S9A, the lens transmits a clear, intense "N", but several fainter images are projected beside the primary image. This "ghosting" effect can be seen in most lenses after moving the mask and refocusing (Figure S9B). Ghost images in the ocelli lenses appear in varying number, size, shape, location, and intensity. One to seven distinct images have been observed through ocelli lenses (Figure S9B). Image doubling is not apparent in ACC lenses (Figure S9C,D).

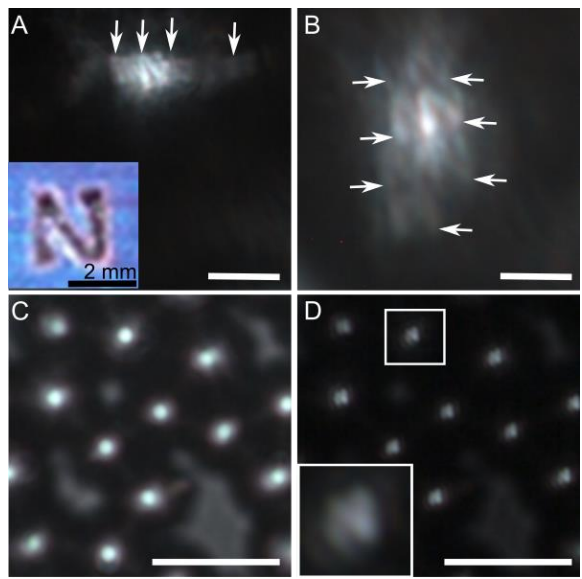

Figure S9. **A)** Ocellus with mask. Inset: "N" mask. **B)** Region from A after the mask was moved and the image was refocused. **C)** Synthetic ACC microlens array without mask. **D)** Synthetic microlens array with mask. Inset: expanded view of boxed region. Scale bars 12.5 µm unless otherwise noted.

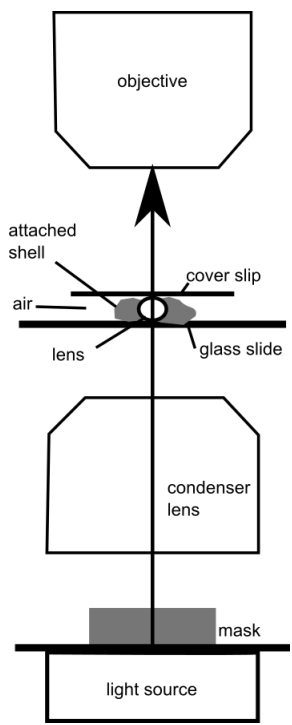

Figure S10. Image transmission apparatus.

# 10 Supplemental References